\documentclass[aps,10pt,superscriptaddress,floatfix,nofootinbib,notitlepage]{revtex4-1}

\usepackage{graphicx}
\usepackage{bm} 
\usepackage{amsmath}
\usepackage{amssymb}
\usepackage{braket}
\usepackage{hyperref}
\usepackage{multirow}
\usepackage{array}
\usepackage{subcaption}
\usepackage{tikz}
\usetikzlibrary{positioning}
\usepackage{float}

\newcounter{subfig}

\begin{document}


\title{A Potential Model Study of the Nucleon's Charge and Mass Radius}

 \author{Daniel Gallimore}
\email{dpgallim@iu.edu}
\affiliation{ Physics Department and Center for Exploration of Energy and Matter,
Indiana University, 2401 N Milo B. Sampson Lane, Bloomington, IN 47408, USA.} 

 \author{Jinfeng Liao}
\email{liaoji@iu.edu}
\affiliation{ Physics Department and Center for Exploration of Energy and Matter,
Indiana University, 2401 N Milo B. Sampson Lane, Bloomington, IN 47408, USA.}

\date{\today}

\begin{abstract} 

We study the charge and mass distributions within a nucleon and compute the associated squared radii based on a potential model approach. Different constituent quark configurations such as $\Delta$, $Y$, and quark-diquark are considered and compared, with model parameters calibrated by experimental measurements of the proton and neutron charge radius. The results suggest that while the charge radius is dictated by quark dynamics, the mass radius is strongly influenced by nonperturbative QCD contributions to a nucleon's mass that are not sensitive to the constituent quarks. As a result, the mass radius could become substantially different from the charge radius. The obtained nucleon mass distributions of different configurations are further used for simulations of the initial conditions in heavy ion collisions. The computed eccentricities $\varepsilon_2$ and $\varepsilon_3$ are found to demonstrate a considerable sensitivity  to the input nucleon profiles, especially to the  mass radius in the peripheral region as well as for systems with fewer participants.


\end{abstract}


\maketitle


\section{Introduction}

The overwhelming majority of mass in our visible universe is accounted for by the nucleons, i.e. protons and neutrons. It is of fundamental significance to understand the mass (or more broadly speaking energy) composition and spatial distributions inside a proton.  One   question attracting substantial recent interests is the mass radius of proton, which is part of the broader quest to understand the gravitational form factors of the proton~\cite{Burkert:2023wzr}.  While the proton charge radius has been studied since the mid-20th century and determined to high precision~\cite{Gao:2021sml,Cui:2022fyr}, its mass radius is still relatively poorly known with notable progress only in the last several years to extract or constrain the value from various experimental measurements and lattice QCD simulations~\cite{GlueX:2019mkq,Mamo2020,Kharzeev2021,Mamo2021,Duran2023,Wang2021,GlueX:2023pev,Polyakov:2018zvc,Burkert:2023atx,Kaiser:2024vbc}.  For example, data for the near-threshold photo-production of $J/\Psi$ at JLab experiments were analyzed to indicate a relatively small mass radius around $ 0.5\rm fm$ which is much smaller than the charge radius of about $0.84 \rm fm$. In heavy ion collision experiments, proton mass radius is potentially the more relevant parameter for the energy smearing width of the binary nucleon interactions which in turn influences the initial conditions of hydrodynamic evolution in these collisions. Comprehensive Bayesian analyses of heavy ion observables appear to prefer considerably larger values, some well above the charge radius \cite{Bernhard2019,JETSCAPE2021,Parkkila2021,Parkkila2022}. The current uncertainty in the mass radius value calls for more investigations  and in particular the possibility of a considerable difference between the mass and charge radius may pose nontrivial constraints on our understanding of partonic structures of the proton. In this work, we aim to provide some insights into this question by calculating the mass distribution of proton and the corresponding mass radius in a potential model approach with constrained charge radius. As the results shall  demonstrate, while the charge radius is mainly controlled by the constituent quarks' wave-function size, the  mass radius is especially sensitive to how the non-perturbative vacuum contributions to the overall energy are treated. Additionally, we will use simulations to demonstrate the impact of proton mass profile on the initial conditions of heavy ion collisions.


\section{Potential Model Setup}

The use of potential models for studying the quark structures of hadrons has a long history with wide applications. Here we adopt the non-relativistic potential model for the proton structure in terms of three constituent quarks. This begins by solving the familiar Schr\"odinger's equation for the proton wave function: 
\begin{equation}\label{eq_Hamiltonian}
    \left(
    \sum_im_i
    +\hat{T}
    +\hat{V} 
    \right)
    \psi (\bm{r}_1,\bm{r}_2,\bm{r}_3)
    =M \, \psi (\bm{r}_1,\bm{r}_2,\bm{r}_3)
\end{equation} 
where $m_i$ and $\bm{r}_i$ ($i=1,2,3$) are the mass and coordinate of each quark while $M$ is the total proton mass. The $\hat{T}$ and 
    $\hat{V} $ operators are the total kinetic and potential energy terms in the Hamiltonian, which we discuss separately below. 
     {It shall be noted that the  non-relativistic description of proton here is an approximation  whiles a rigorous treatment of proton structures (e.g. in terms of form factors) would necessarily require relativistic dynamics~\cite{Miller:2002ig,Smith:2004dn,Miller:2009sg,Cloet:2012cy,Perdrisat:2006hj,Punjabi:2015bba,Djukanovic:2021qxp}.  }

\subsection{The Kinetic Term}

Clearly the problem at hand is a general 3-body problem, with 9 degrees of freedom in the coordinate space. It is well known that one could choose the center-of-mass (CM) frame and use the following set of Jacobi coordinates to simply the kinematics: 
\begin{align}
    \bm{x}_1
    &=\frac{2}{\sqrt{3}}
    \left(
    \bm{r}_3-\frac{\bm{r}_1+\bm{r}_2}{2}
    \right), \\
    \bm{x}_2
    &=\bm{r}_2-\bm{r}_1, \\
    \bm{x}_3
    &=\frac{\bm{r}_1+\bm{r}_2+\bm{r}_3}{3}.
\end{align}
The kinetic energy operator in the CM frame is then 
\begin{equation}
    \hat{T}
    =-\frac{1}{m}
    \left(
    \nabla_{x_1}^2
    +\nabla_{x_2}^2
    \right)
\end{equation}
where for simplicity we assume all quark masses are equal, i.e. $m_i\equiv m$. 

Clearly the internal Jacobi coordinates together span a 6-dimensional Euclidian space $(\bm{x}_1,\bm{x}_2)$. In analogy to the usual spherical coordinate system in 3-dimension, one can also cast the $(\bm{x}_1,\bm{x}_2)$ space into a coordinate system of one hyper-radius $\rho$ and a set of five angles $\bm{\Omega}$~\cite{Marcucci2020},  where
\begin{align}
    \rho&=\sqrt{x_1^2+x_2^2}, \\
    \bm{\Omega}&=(\theta_1,\phi_1,\theta_2,\phi_2,\varphi).
\end{align}
Four of the angles are just the polar and azimuthal angles of the internal Jacobi coordinates, and one is the hyperangle $\varphi$, defined by
\begin{equation}
    \cos\varphi=\frac{x_2}{\sqrt{x_1^2+x_2^2}},\quad0\leq\varphi\leq\pi/2.
\end{equation}
In these coordinates, the kinetic energy operator becomes
\begin{equation}
    \hat{T} 
    =-\frac{1}{m}
    \left(
    \frac{\partial^2}{\partial\rho^2}
    +\frac{5}{\rho} 
    \frac{\partial}{\partial\rho} 
    -\frac{\Lambda^2(\bm{\Omega})}{\rho^2}
    \right),
\end{equation}
where $\Lambda^2(\bm{\Omega})$ is the so-called grand angular momentum.

Again in analogy to the spherical harmonics in 3-dimension, one can expand a general wave function in terms of a convenient basis: the eigenfunctions of $\Lambda^2(\bm{\Omega})$ which are the  hyper-spherical harmonics $\mathcal{Y}(\bm{\Omega})$  with corresponding eigenvalues $K(K+4)$, where $K\geq0$ is the grand angular momentum quantum number~\cite{Marcucci2020}.  For the purpose of this study,  we focus solely on the $s$-wave modes. In this case (i.e. when the total orbital angular momentum associated with the Jacobi coordinates $(\bm{x}_1,\bm{x}_2)$ is zero, $\bm{x}_1$ and $\bm{x}_2$ must individually have the same orbital angular momentum quantum number $l$. The allowed grand angular momenta are
\begin{equation}
    K=2(l+n),
\end{equation} 
where $n\geq0$ is a hyper-radial quantum number. The corresponding $s$-wave harmonics are
\begin{equation}
    \mathcal{Y}_{ln}(\omega,\varphi)
    =(-1)^l
    \frac{\sqrt{2l+1}}{4\pi} 
    P_l(\cos\omega)
    \mathcal{P}_n^{l,l}(\varphi),
\end{equation}
with $\bm{x}_1\cdot\bm{x}_2=x_1x_2\cos\omega$ and $P_l$ is a Legendre polynomial. The special function $\mathcal{P}_n^{l,l}$ is defined as
\begin{equation*}
    \mathcal{P}_n^{l,l}(\varphi)
    =\frac{2\sqrt{n!(l+n+1)\Gamma(2l+n+2)}}{\Gamma(l+n+3/2)}
    \sin^l\varphi
    \cos^l\varphi
    P_n^{l+1/2,l+1/2}(\cos2\varphi),
\end{equation*}
where $P_n^{l+1/2,l+1/2}$ is a Jacobi polynomial. The harmonics are orthogonal via
\begin{equation}
    \int d\bm{\Omega}\,
    \mathcal{Y}_{ln}(\bm{\Omega})
    \mathcal{Y}_{l'n'}(\bm{\Omega})
    =\delta_{ll'}
    \delta_{nn'},
\end{equation}
with $d\bm{\Omega}=(8\pi^2\sin\omega\,d\omega)(\cos^2\varphi\sin^2\varphi\,d\varphi)$.


\subsection{The Potential Term}

Let us now discuss the potential term. A natural starting point for potential models is the quark-antiquark potential (in the color singlet channel) that has been comprehensively studied both via lattice gauge theory calculations and via heavy quarkonium spectroscopy studies within potential models. A popular parametrization of such a 2-body potential is the Cornell potential below: 
\begin{equation}\label{eq_Cornell}
    V(\bm{r}_i,\bm{r}_j)=-\frac{\kappa}{|\bm{r}_i-\bm{r}_j|}+\sigma|\bm{r}_i-\bm{r}_j|-V_0.
\end{equation}
The Coulomb term accounts for one-gluon exchange at short distance and the linear term confines the quarks at long distance~\cite{Richard1992}. The Coulomb and string tension parameters, $\kappa$ and $\sigma$, are flavor-independent and can be fixed as $\kappa=0.3$ and $\sigma=0.18\,\text{GeV}^2$, which are consistent with lattice simulations of heavy quarkonia \cite{Bali2001,Mateu2019}. The term $V_0$ is a constant offset.

Now for the proton structure, one needs to model the 3-body potential for the three constituent quarks. There are different choices, such as the so-called $\Delta$ and $Y$ models as well as other models (e.g. diquark clustering model to be discussed later). The $\Delta$ model is a straightforward generalization of 2-body potential into a simple summation of pair-wise potentials between each pair:
\begin{equation}\label{eq_Delta}
    V_\Delta(\bm{r}_1,\bm{r}_2,\bm{r}_3)
    =-\frac{\kappa}{2}
    \sum_{i<j}
    \frac{1}{|\bm{r}_i-\bm{r}_j|}
    +\frac{\sigma}{2}
    \sum_{i<j}
    |\bm{r}_i-\bm{r}_j|
    -\frac{3}{2}V_0.
\end{equation}
The factor of $1/2$ multiplying the Coulomb and confining terms reflects the fact that each pairwise potential extends between two quarks instead of between a quark and an antiquark \cite{Richard1992}. The so-called $Y$ model, on the other hand, accounts for genuine 3-body force beyond just pair-wise interactions by postulating a central junction to which each constituent quark is connected by chromoelectric flux tubes carrying the nonperturbative confining energies~\cite{Richard1992}. This model also includes the perturbative short-range Coulomb interactions in a pair-wise way.  The total potential, therefore, takes the following form: 
\begin{equation}
    V_Y(\bm{r}_1,\bm{r}_2,\bm{r}_3)
    =-\frac{\kappa}{2}
    \sum_{i<j}
    \frac{1}{|\bm{r}_i-\bm{r}_j|}
    +\sigma \left [ \min_{\bm{x}}\sum_i|\bm{r}_i-\bm{x}| \right ]
    -\frac{3}{2}V_0.
\end{equation} 
For any given 3-quark configuration, the minimization of the junction term requires $\bm{x}=\bm{x}_\text{F}$, the so-called Fermat point. It may be tempting to ask how different the $\Delta$ and $Y$ potentials would be. One can show that \cite{Richard1992}
\begin{equation}
    \frac{1}{2}\sum_{i<j}|\bm{r}_i-\bm{r}_j|
    \leq
    \sum_i|\bm{r}_i-\bm{x}_\text{F}|
    \leq
    \frac{1}{\sqrt{3}}\sum_{i<j}|\bm{r}_i-\bm{r}_j|,
\end{equation}
so the $\Delta$ and $Y$ models will often produce similar quantitative results. 

Using the method in \cite{Carlson1983}, $\bm{x}_\text{F}$ can be expressed purely in terms of the quark separations $r_{ij}\equiv|\bm{r}_j-\bm{r}_i|$. Take the quarks to be the vertices of a triangle. If one of the interior angles of the triangle is greater than $120^\circ$, then $\bm{x}_\text{F}$ coincides with the quark at that vertex. 
\begin{equation}
    |\bm{r}_i-\bm{x}_\text{F}|
    =\frac{B^2-2r_{jk}^2}{2A}
\end{equation}
holds cyclically for $(i,j,k)$, where
\begin{align}
    A^2&=\frac{3}{2}\left(B^2-\rho^2\right), \\
    B^2&=\frac{3}{2}\left(\rho^2+\sqrt{\frac{\eta^4-\rho^4}{3}}\right), \\
    \rho^2&=\frac{2}{3}\left(r_{12}^2+r_{23}^2+r_{31}^2\right), \\
    \eta^4&=\frac{16}{9}\left(r_{12}^2r_{23}^2+r_{23}^2r_{31}^2+r_{31}^2r_{12}^2\right).
\end{align}
The quantity $\rho$ is the very same hyperradius from the previous section. It is simple to show that
\begin{equation}
    \sum_i|\bm{r}_i-\bm{x}_\text{F}|
    =A.
\end{equation}
For three equal mass quarks, $\bm{x}_\text{F}$ is often close to the center-of-mass, and it is only when using a quark-diquark geometry that the two may be very different. Thus, we use the simpler potential
\begin{equation}\label{eq_Y}
    V_Y(\bm{r}_1,\bm{r}_2,\bm{r}_3)
    =-\frac{\kappa}{2}
    \sum_{i<j}
    \frac{1}{|\bm{r}_i-\bm{r}_j|}
    +\sigma\sum_ir_i
    -\frac{3}{2}V_0
\end{equation} 
in practice (with the center-of-mass fixed at the origin). Numerically, the proton charge radius calculated in this simplified $Y$ model differs from the more rigorous potential by less than $1\%$.

For each of the above models, there are still two parameters: the offset constant $V_0$ as well as the constituent quark mass $m$. The former will be simply fixed by equating the ground state energy and the physical proton mass $M=0.938\,\text{GeV}$. The latter will be fixed by reproducing the proton charge radius at $0.84\,\text{fm}$.

\subsection{Solving the Radial Equations}

Given both kinetic and potential terms, the next step is to find solutions to eq. (\ref{eq_Hamiltonian}) of the form
\begin{equation}
    \psi(\rho,\omega,\varphi)
    =\sum_{l,n}
    \frac{u_{ln}(\rho)}{\rho^{5/2}}
    \mathcal{Y}_{ln}(\omega,\varphi),
\end{equation}
where $u_{ln}(\rho)$ is a hyper-radial wave function. Then
\begin{equation}
    (M-3m)u_{ln}
    =-\frac{1}{m}
    \left(
    \frac{d^2}{d\rho^2}
    -\frac{(K+3/2)(K+5/2)}{\rho^2}
    \right)
    u_{ln}
    +\sum_{l'n'}
    \left(
    \int d\bm{\Omega}
    \mathcal{Y}_{ln}V\mathcal{Y}_{l'n'}
    \right)
    u_{l'n'},
\end{equation}
which can be written in the dimensionless form
\begin{equation}\label{eq_dimless}
    \mathcal{E}u_{ln}(\xi)
    =-u_{ln}''(\xi)
    +\frac{(K+3/2)(K+5/2)}{\xi^2}u_{ln}(\xi)
    +\frac{\lambda}{2}
    \sum_{l'n'}
    \left(
    -\frac{\alpha_{ln}^{l'n'}}{\xi} 
    +\beta_{ln}^{l'n'}\xi
    \right)
    u_{l'n'}(\xi)
\end{equation}
with the definitions $\gamma\equiv\sqrt{\kappa/\sigma}$, $\xi\equiv\rho/\gamma$, $\lambda\equiv m\gamma\kappa$, and $\mathcal{E}\equiv m\gamma^2(M-3m+3V_0/2)$. The quantities $\alpha_{ln}^{l'n'}$ and $\beta_{ln}^{l'n'}$ are integrals over $\bm{\Omega}$ that depend on the form of the potential and quantify the mixing of different hyper-spherical modes due to the interaction.   Truncating at appropriate $l$ and $n$, eq. (\ref{eq_dimless}) can be numerically solved using the inverse power method for coupled equations \cite{Crater1994}. For the ground state that we are interested in,  it can be numerically verified that $u_{00}$ is many orders of magnitude more significant than any other $u_{ln}$, so for all practical purposes it is safe to assume $\psi\equiv\psi(\rho)$. Note that we only focus on the spatial wave function and assume the spin, color and flavor degrees of freedom to be in the usual quark model configurations for baryon octet. Specifically, the spin-isospin structure for a proton should be such that $| p_\uparrow \rangle  = \frac{1}{\sqrt{18}} \left[ 2 u_\uparrow u_\uparrow d_\downarrow -u_\uparrow u_\downarrow d_\uparrow - u_\downarrow u_\uparrow d_\uparrow - u_\uparrow d_\uparrow u_\downarrow + 2 u_\uparrow d_\downarrow u_\uparrow   - u_\uparrow
 d_\uparrow u_\downarrow  - d_\uparrow u_\uparrow u_\downarrow - d_\uparrow u_\downarrow u_\uparrow + 2 d_\downarrow u_\uparrow u_\uparrow   \right]   $~\cite{Peskin:2019iig}, where the $\uparrow$ or $\downarrow$ indicates spin up or down state~\cite{Peskin:2019iig}. A similar structure can be written down for a neutron by swapping the $u$ and $d$ flavors of quarks. Once the wave function is obtained, one can compute various properties of the proton such as the charge radius and mass radius, which we shall discuss next. 


\section{Proton Charge Radius}


As a first (and mandatory) ``sanity check," one would like to calibrate the model with the proton charge radius which is  experimentally known to high precision. Naively, one would use the wave function from solving the Schr\"odinger equation to evaluate the following charge density distribution: 
\begin{equation} \label{eq_rho_e}
    \rho_e(\bm{r})
    =  \int \prod_{j=1,2,3}\!d^3r'_{j}\ 
   \psi^\dag(\bm{r}'_{k=1,2,3})
 \left [ \sum_{i=1,2,3} q_i    \delta(\bm{r}-\bm{r}'_i)  \right]
    \psi(\bm{r}'_{k=1,2,3}) = \sum_{i=1,2,3} q_i  \int \prod_{j=1,2,3}\!d^3r'_{j}\ 
    \psi^\dag  \psi \ 
    \delta(\bm{r}-\bm{r}'_i),
\end{equation}
where the sum is over three valence quarks with flavor $u$, $u$, $d$ and electric charge $q_i=2/3,2/3, -1/3$ respectively. Then one could compute the charge radius via the second moment of the above distribution, i.e.: 
\begin{equation} \label{eq_Re_1} 
R_e^2 \equiv \int d^3r\,  \rho_e(\bm{r}) r^2 \,\, ,
\end{equation}
where the normalization $\int d^3r\,   \rho_e(\bm{r}) = 1$ is implied for the total charge of the proton. 
 {Let us emphasize that the above definition of charge radius is based on the non-relativistic description in this study. In a more rigorous relativistic approach, the so-called charge radius shall be extracted from the slope of the electromagnetic form factors at low momentum transfer~\cite{Miller:2002ig,Smith:2004dn,Miller:2009sg,Cloet:2012cy,Perdrisat:2006hj,Punjabi:2015bba,Djukanovic:2021qxp}. Nevertheless, our goal here is to investigate within the same non-relativistic potential model how much the charge distribution and mass distribution could become different.}

Clearly in this naive approach the charge radius will be close to the wave function size. However, neither model, $\Delta$ or $Y$, is able to produce the correct proton charge radius of about $0.84\,\text{fm}$ using the above recipe and realistic potential model parameters. 
With the usually adopted physical value for the string tension $\sigma$, the wave function becomes compressed around the center-of-mass. At values of $\sigma$ consistent with lattice simulations, the constituent quarks are confined in a region of size well below $0.84\,\text{fm}$, along with the charges they carry. Consequently, the naive calculation gives only about half the expected radius. This is actually a well known issue, rooted in the non-relativistic approximation. 

Quarks, like all spin-1/2 particles, are ultimately described by Dirac fields. However, in its native representation, the Dirac Hamiltonian is not energy-separating and its dynamical variables do not in general satisfy the Poincar\'e algebra. This is in direct conflict with Schr\"odinger quantum mechanics, whose Hamiltonian does not mix particle and antiparticle states and whose dynamical variables do satisfy the Poincar\'e algebra. Before taking its nonrelativistic limit, the Dirac theory must be written in a representation whose Hamiltonian and dynamical variables satisfy the same properties as the Schr\"odinger theory. The appropriate representation was first proposed by Foldy and Wouthuysen \cite{Foldy1950}, with a later correction by Eriksen \cite{Eriksen1960}. However, this transformation causes an issue when calculating the charge radius of a system of point-like objects using the resulting nonrelativistic wave function. Namely, the squared charge radius should not be computed via the naive formula as $\sum_iq_i\braket{r_i^2}$. There are various ways of addressing this issue in practice \cite{Stanley1980,Hayne1982,Godfrey1985,Semay1997,Lombard2000,Guo2020}. The problem arises because localized states in the Dirac theory are no longer localized after the Foldy-Wouthuysen (FW) transformation. Instead, they become spread over an intrinsic radius $\sim1/m$, a phenomenon known as \textit{zitterbewegung}. In contrast with their heavier cousins, light quarks cannot be treated as localized particles inside baryons, even in the nonrelativistic picture.

We now implement this necessary correction. The single-particle density function $\rho_\text{SP}$ written in terms of the Schr\"odinger wave function $\psi$ is
\begin{equation}
    \rho_\text{SP} (\bm{r})
    =\int d^3r'\,
    \psi^\dag(\bm{r}')
    U_{FW}\delta(\bm{r}-\bm{r}')U^\dag_{FW}
    \psi(\bm{r}')
\end{equation}
since $\psi$ is essentially the FW transformed Dirac wave function in the nonrelativistic limit. We use a suitable approximation for $U_{FW}\delta(\bm{r}-\bm{r}')U^\dag_{FW}$ based on the work of Eriksen \cite{Eriksen1960}. The classical limit of the charge density of a stationary, localized Dirac particle is approximately the product of the charge and the normalized distribution
\begin{equation}
    D(\bm{s})
    =\frac{1}{\sqrt{2}}\delta(\bm{s})
    +\mathcal{W}(s),
\end{equation}
with
\begin{equation}
    \mathcal{W}(s)
    =\frac{\sqrt{2}}{8\pi^2s}
    \frac{d}{ds}
    \frac{1}{s}
    \left(
    \Gamma\!\left(\frac{5}{4}\right)
    W_{-\frac{1}{4},\frac{1}{2}}(2ms)
    -\Gamma\!\left(\frac{3}{4}\right)
    W_{\frac{1}{4},\frac{1}{2}}(2ms)
    \right).
\end{equation}
The special functions $W_{\alpha\beta}$ are Whittaker functions. We define the single-particle density function using $D$ rather than $\delta$, i.e., 
\begin{equation}
    \rho_\text{SP}(\bm{r})
    =\int d^3r'
    \,\psi^\dag(\bm{r}')
    D(\bm{r}-\bm{r}')
    \psi(\bm{r}').
\end{equation} 
Such a correction produces an extended quark with a relativistic smeared radius $\sim1/m$ whose charge density exists well beyond its mean position. For example, a light quark with mass $270\,\text{MeV}$ has an effective radius of about $0.7\,\text{fm}$, which is comparable to the proton charge radius. This smeared quark profile originates from the FW transform, and is a necessary component in any light quark potential model.

\begin{figure}[!htb]
    \centering
    \includegraphics[width=0.6\textwidth]{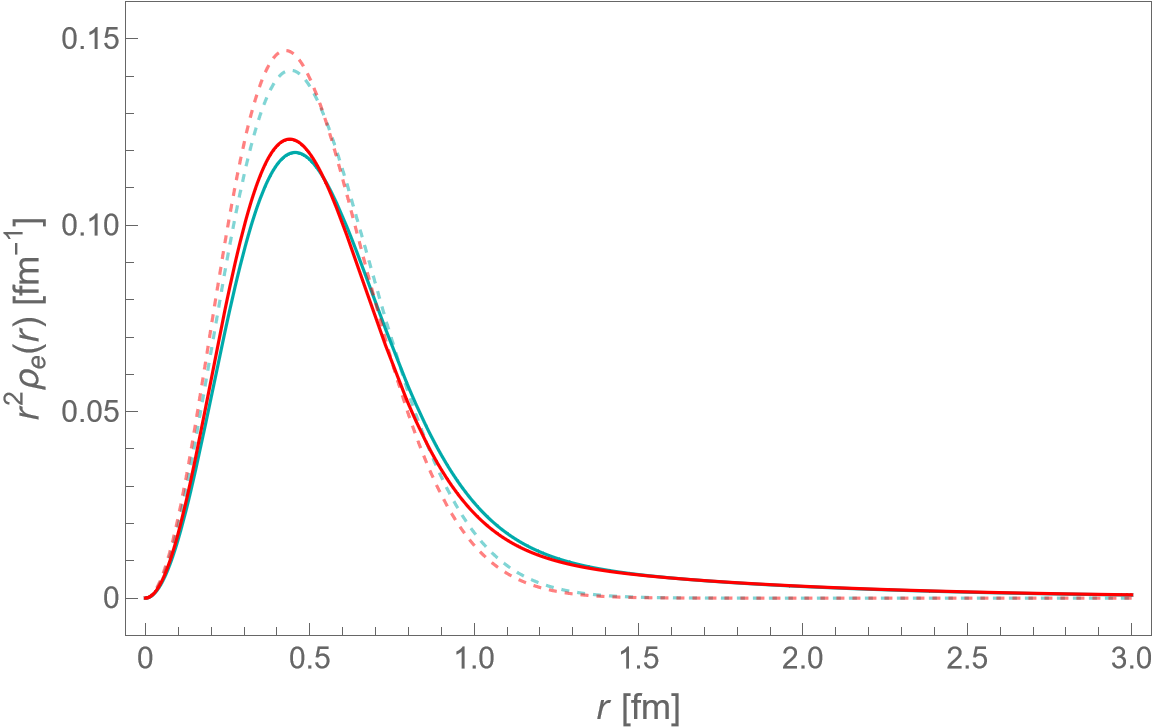}
    \caption{$\Delta$ (cyan) and $Y$ (red) proton charge distributions. The substitution $\delta\rightarrow D_i$ smears the proton charge density and increases the charge radius. The dashed curves correspond to point-like quarks and the solid curves incorporate smearing.}
    \label{fig_rho_e}
\end{figure}

Based on this FW recipe, we write down the modified charge density of proton: 
\begin{equation}\label{eq_rho_e}
    \rho_e(\bm{r})
    =  \sum_{i=1,2,3} q_i  \int \prod_{j=1,2,3}\!d^3r'_{j}\ 
   \psi^\dag(\bm{r}'_{k=1,2,3})
   D_i(\bm{r}-\bm{r}'_i)   
    \psi(\bm{r}'_{k=1,2,3}) .
\end{equation}
(The subscript on the $D_i$ indicates that the charges may in general have different masses.) The total proton charge density for the $\Delta$ and $Y$ models is illustrated in fig. \ref{fig_rho_e}. Combining eqs. (\ref{eq_Re_1}) and (\ref{eq_rho_e}), and substituting $\bm{s}_i=\bm{r}-\bm{r}_i$, gives
\begin{equation}\label{eq_qrad}
    R_e^2
    =\sum_iq_i\braket{r_i^2}
    +\sum_iq_i
    \int d^3s_i\,s_i^2
    \mathcal{W}_i(s_i)
\end{equation}
after expanding and simplifying the integral. In the $\Delta$ and $Y$ models, $m_i\equiv m$ and $\sum_iq_i=1$ for the proton, so the above equation implies a squared proton charge radius that is larger than $\sum_iq_i\braket{r_i^2}$. We then use the expected charge radius $R_e=0.84\ \rm fm$ to constrain our model parameters: they are $m=0.2744\,\text{GeV}$, $V_0=1.103\,\text{GeV}$ and $m=0.2668\,\text{GeV}$, $V_0=1.225\,\text{GeV}$ for the $\Delta$ and $Y$ models, respectively.

\section{Proton Mass Radius}

We now turn to the calculation of the energy density distribution in the center-of-mass (i.e., rest) frame of the proton, which can then be further used to compute the proton mass radius. 
 {We start by emphasizing that the same proton wavefunction, obtained from the potential model introduced in Sec. II with parameters fixed by charge radius in Sec. III, will be used to compute the energy distribution in this section.} Whereas the charges used to calculate the charge radius are restricted to the quarks themselves, the distribution of energy that determines the mass radius depends on the nature of the energy contents. 

 {
To make this point clear, let us first try to introduce an energy density distribution $\rho(\bm{r})$ in analogy to the charge distribution defined in eq.(\ref{eq_rho_e}:  
\begin{equation} \label{eq_rho_m}
    \rho(\bm{r})
    =  \int \prod_{j=1,2,3}\!d^3r'_{j}\ 
   \psi^\dag(\bm{r}'_{k=1,2,3}) \, 
 \hat{h}(\bm{r} | \bm{r}'_{k=1,2,3} ) \, 
    \psi(\bm{r}'_{k=1,2,3}) 
\end{equation} 
Different from the charge density operator in eq.(\ref{eq_rho_e}, it is not obvious how the local energy density operator $\hat{h}(\bm{r})$ should be defined. Given that the total energy consists of kinetic, potential and offset energy contributions, it is natural to consider $\rho(\bm{r})$ as a sum over spatial distributions of the  kinetic, potential and offset energies respectively. Technically speaking, for each given configuration (which means a set of coordinate $\bm{r}'_{i=1,2,3}$ for the constituent quarks), one needs to come up with a spatial distribution of kinetic, potential and offset energies. Then a convolution with the proton wavefunction (i.e. probability distribution for each configuration) gives the averaged energy density distribution.
}

 {
Once the energy density distribution $\rho(r)$ is known, one can similarly define a RMS mass radius analogously to the charge radius in eq.(\ref{eq_Re_1}): 
\begin{eqnarray}
R_m^2 = \frac{\int d^3r \, \rho(r) \, r^2 }{\int d^3r \, \rho(r)} = \frac{\int d^3r \, \rho(r) \, r^2}{M}
\end{eqnarray}
We next discuss the computation of kinetic, potential and offset terms respectively. 
}


\subsection{ {Spatial Distribution of Kinetic Energy}}


We begin by calculating the kinetic contributions associated with quarks to the total energy density. The mass energy density of the $i$th quark  is simply $m_iD_i(\bm{r}-\bm{r}_i)$. We use the same extended quark profile $D_i$ as the previous section since a quark---even one with an extended profile---has no internal structure that would cause its charge radius to differ from its mass radius. Our choice of kinetic energy density must be made more carefully since there is no unique kinetic energy density operator \cite{Muga2005,Krogel2013,Maier2019}. Since we want the moments of the expectation value to be positive, we choose the positive-definite form
\begin{equation}
    \tau_i
    =\frac{\bm{p}_{r_i}\cdot D_i(\bm{r}-\bm{r}_i)\ \bm{p}_{r_i}}{2m_i}
\end{equation}
to be the kinetic energy density of the $i$th quark. 

In the $\Delta$ and $Y$ models where the quark masses are equal, 
\begin{equation}
    \braket{\tau_i}
    =\frac{1}{2m}
    \int d^3x_1d^3x_2
    \,D(\bm{r}-\bm{r}_i)\,
    |\bm{\nabla}_{r_i}\psi|^2.
\end{equation}
Due to symmetry, $\braket{\tau_1}=\braket{\tau_2}=\braket{\tau_3}$, so only one of these needs to be calculated. In terms of the internal coordinates,
\begin{align}
    \bm{r}_1
    &=-\frac{1}{2}
    \left(
    \frac{1}{\sqrt{3}}\bm{x}_1+\bm{x}_2
    \right), \\
    \bm{r}_2
    &=-\frac{1}{2}
    \left(
    \frac{1}{\sqrt{3}}\bm{x}_1-\bm{x}_2
    \right), \\
    \bm{r}_3
    &=\frac{1}{\sqrt{3}}\bm{x}_1.
\end{align}
Additionally,
\begin{align}
    \bm{p}_{r_1}
    &=-\frac{1}{\sqrt{3}}\bm{p}_{x_1}
    -\bm{p}_{x_2}, \\
    \bm{p}_{r_2}
    &=-\frac{1}{\sqrt{3}}\bm{p}_{x_1}
    +\bm{p}_{x_2}, \\
    \bm{p}_{r_3}
    &=\frac{2}{\sqrt{3}}\bm{p}_{x_1}.
\end{align}
With $\braket{\tau_i}$ written in terms of internal variables, the calculation is straightforward.

\subsection{  {Spatial Distribution of Potential Energy}}

The Coulomb potential and the confining linear potential energies are not localized at the quarks themselves, but are carried by chromoelectric fields, which we model as uniform tubes of flux. We introduce the coordinates
\begin{align}
    \bm{r}_{ij}&=\bm{r}_j-\bm{r}_i, \\
    \bm{s}_{ij}&=\bm{r}-\frac{\bm{r}_i+\bm{r}_j}{2}
\end{align}
to simplify the following expressions. The Coulomb energy density between a quark and antiquark (or quark and diquark) located at $\bm{r}_i$ and $\bm{r}_j$ is assumed to take the form: 
\begin{equation}
    v_{ij}^\text{Coul}
    =-\frac{\kappa}{\pi a^2r_{ij}^2}
    \,\Theta\!\left(
    1-\frac{4(\bm{r}_{ij}\cdot\bm{s}_{ij})^2}{r_{ij}^4}
    \right)
    \Theta\!\left(
    1-\frac{s_{ij}^2}{a^2}+\frac{(\bm{r}_{ij}\cdot\bm{s}_{ij})^2}{a^2r_{ij}^2}
    \right),
\end{equation}
with $\Theta$ the unit step function. The constant $a$ is the perpendicular radius of the flux tube distribution. The confining potential energy is likewise distributed as
\begin{equation}\label{eq_conf}
    v_{ij}^\text{Conf}
    =\frac{\sigma}{\pi a^2}
    \,\Theta\!\left(
    1-\frac{4(\bm{r}_{ij}\cdot\bm{s}_{ij})^2}{r_{ij}^4}
    \right)
    \Theta\!\left(
    1-\frac{s_{ij}^2}{a^2}+\frac{(\bm{r}_{ij}\cdot\bm{s}_{ij})^2}{a^2r_{ij}^2}
    \right),
\end{equation}
In the $\Delta$ model, $\bm{r}_i$ and $\bm{r}_j$ are the positions of two quarks, so a factor of $1/2$ needs to be included in the definitions of $v_{ij}^\text{Coul}$ and $v_{ij}^\text{Conf}$. Additionally, in the Y model, the confining energy lies between all three quarks rather than between each pair, so $\bm{r}_i$ should be interpretted as the location of the $i$th quark and $\bm{r}_j$ as the location of the central junction in $v_{ij}^\text{Conf}$.

We set $a=0.1\,\text{fm}$; though, any value between $0.1$ and $0.4\,\text{fm}$ will produce similar mass radii. We also use the same value of $a$ in both $v_{ij}^\text{Coul}$ and $v_{ij}^\text{Conf}$ since using a different value in the range $0.1$ and $0.4\,\text{fm}$ does not seem to produce substantial differences in the resulting mass radii.

\subsection{ {Spatial Distribution of Offset Energy}}

The offset energy is the most mysterious contribution to the potential. The key feature of this contribution is that it is the energy of a proton that is {\em insensitive} to where the constituent quarks are. This energy could originate from QCD vacuum contributions, nonperturbative gluonic field energies, etc. It could also account for discrepancies in the constituent quark model, from relativistic corrections to the dynamical energy \cite{Jaczko1998}, to spin-spin, spin-orbit, and tensor corrections to the potential \cite{Mateu2019}, and likely more. 

\begin{figure}[!htb]
    \centering
    \includegraphics[width=0.6\textwidth]{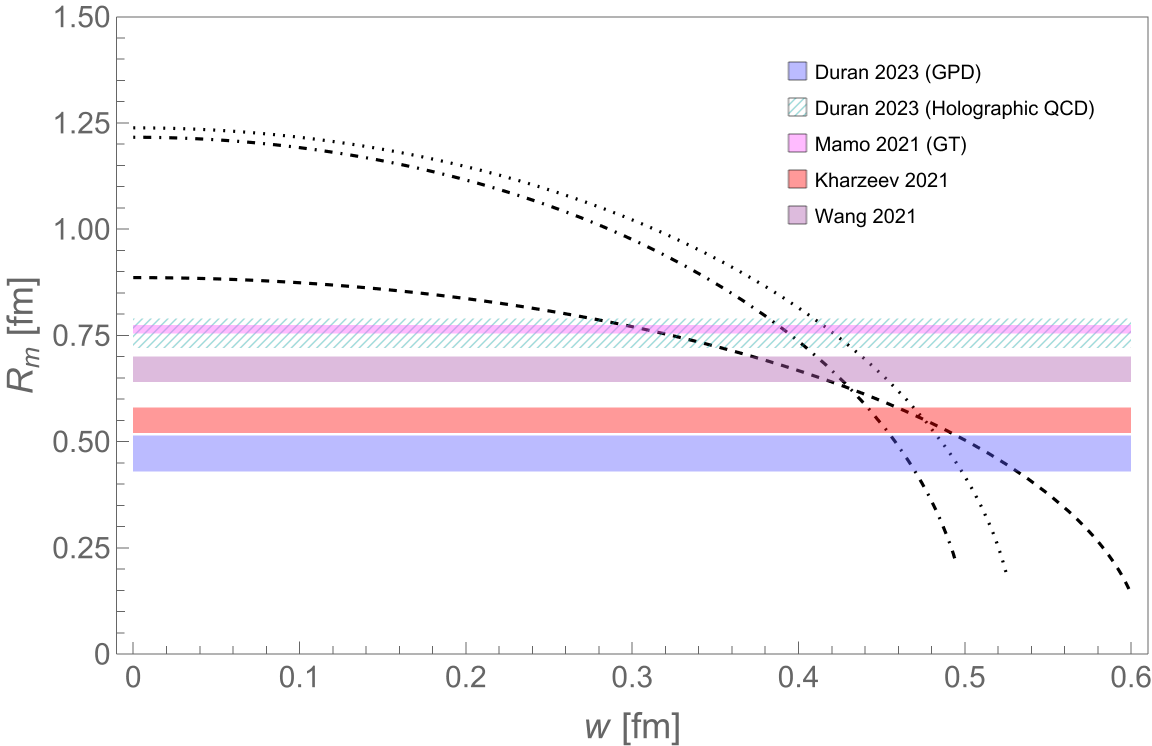}
    \caption{Mass radius $R_m$ versus offset energy width $w$ in the $\Delta$ (dotted),  $Y$ (dot-dashed), and diquark (dashed) models. Colored bands represent mass radius values extracted from various previous model analyses or experimental measurements~\cite{Kharzeev2021,Mamo2021,Duran2023,Wang2021}.}
    \label{fig_radius_vs_width}
\end{figure}

For the purpose of computing the proton energy density profile and the associated mass radius, what perhaps matters the most is the fact that the offset energy is associated with the entire proton rather than with any of the individual quarks. Therefore, we will assume the offset energy is distributed with respect to the center-of-mass in a Gaussian form: 
\begin{equation} \label{eq_offset_dist}
  v_{\rm offset} =  -\frac{V_0}{(2\pi w^2)^{3/2}}
    \exp\!\left(-\frac{r^2}{2w^2}\right)
\end{equation} 
for a quark and antiquark (or quark and diquark). A factor of $3/2$ is included in the $\Delta$ and $Y$ models. Here, the Gaussian width parameter $w$ characterizes the scale (in $\rm fm$) of the spatial spread of this energy. As we shall demonstrate later with numerical results, the offset term has a significant impact on the proton energy density distribution and its mass radius. The relationship between mass radius and the width parameter $w$ is shown in fig. \ref{fig_radius_vs_width}. 

\subsection{Energy Density Distribution and Mass Radius}

\begin{figure}[!htb]

\setcounter{subfig}{0}
  \newcommand\typecap{\stepcounter{subfig}\captiontext*[\value{subfig}]{}}
  \captionsetup{position=bottom, skip=3pt}
  \centering
  \begin{subcaptiongroup}
    \captionlistentry{}\label{fig_mass_dists_Delta}   
    \captionlistentry{}\label{fig_mass_dists_Y}
    \tikz[
      node distance=3pt,
      caption/.style={
        anchor=north east,
        font=\Large,
        outer sep=10pt,
      },
      relpos/.style={
        right=of #1.south east,
        anchor=south west,
      },
    ] {
      \node (A) {\includegraphics[width=0.48\textwidth]{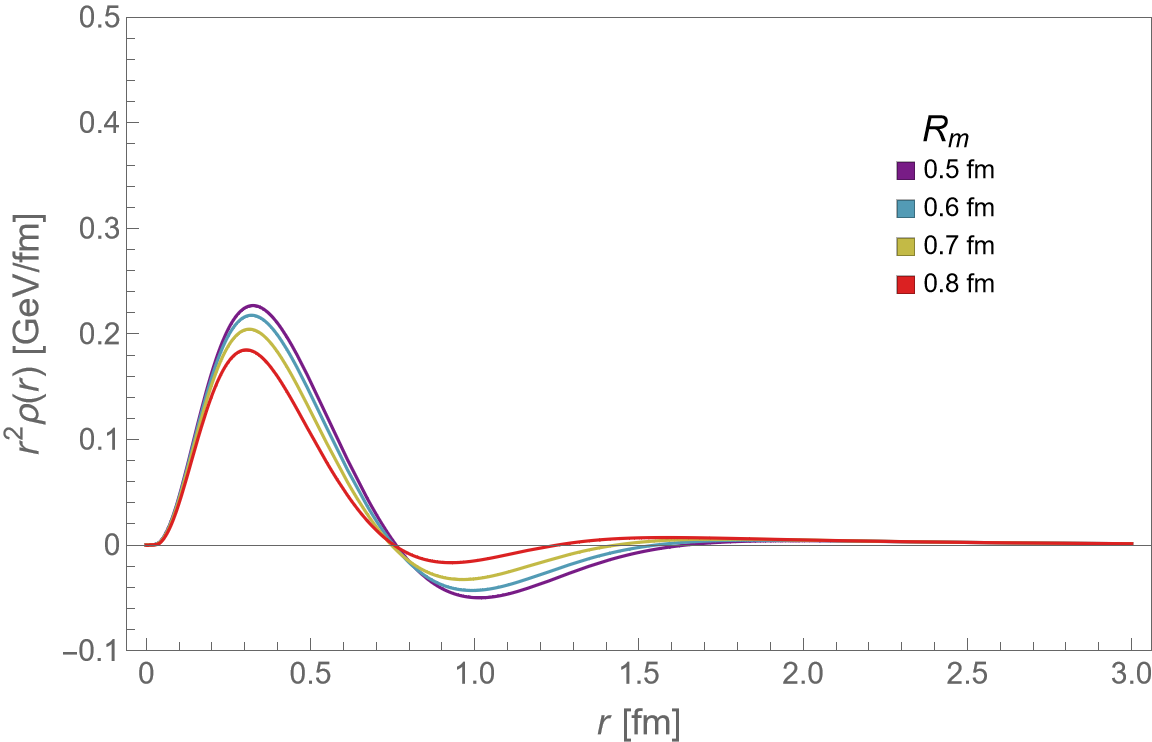}};
      \node (B) [relpos=A] {\includegraphics[width=0.48\textwidth]{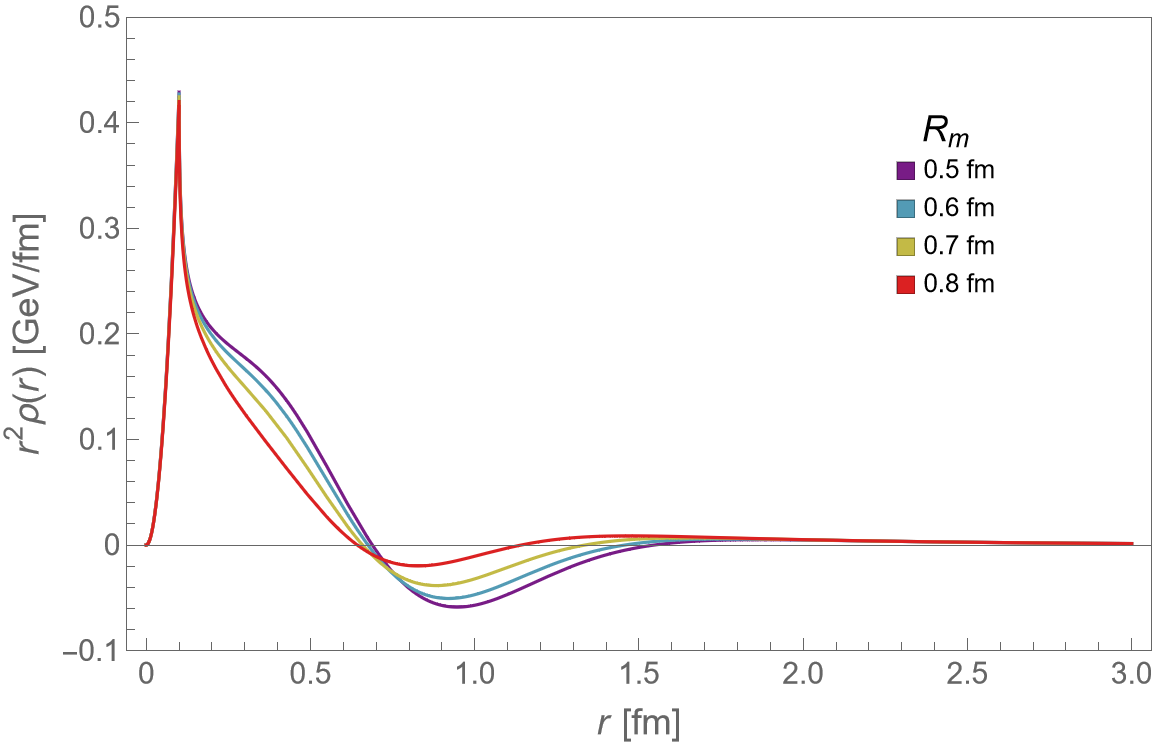}};
      \node at (A.north east) [caption] {\typecap};
      \node at (B.north east) [caption] {\typecap};
    }
  \end{subcaptiongroup}
  \caption{Proton mass distributions in the $\Delta$ model (\subref{fig_mass_dists_Delta}) 
 and $Y$ model (\subref{fig_mass_dists_Y}) corresponding to different choices of mass radius $R_m$.}
   \label{fig_mass_dists}

\end{figure}

Adding all the components together, one can 
then obtain the energy density distribution  $\rho(r)$ by convolution with the proton wave function. The results, plotted in terms of $r^2\rho(r)$, for the $\Delta$ and $Y$ models are shown in fig. \ref{fig_mass_dists}. Different curves are obtained from different choices of the offset energy Gaussian width parameter, labeled by the corresponding mass radius value. As one can see, the energy distribution of both models contains a prominent positive crest close to the center-of-mass and a minor negative trough at larger distances. That being said, the two models differ from each other in the prominence of the short-distance peak, with a rather sharp structure appearing for the $Y$ model due to the junction. 

As already mentioned, the mass radius is primarily sensitive to the Gaussian width of the offset term. Indeed, as clearly shown in fig.~\ref{fig_radius_vs_width}, the computed RMS mass radius rapidly decreases with increasing the width parameter $w$ for both $\Delta$ and $Y$ models, only demonstrating a relatively mild quantitative difference between the two models. This observation could potentially offer an opportunity to constrain the Gaussian width parameter $w$ if the proton mass radius could be experimentally determined with high accuracy through certain measurement as has been done for the proton charge radius.

\section{Neutron Structure and The Diquark Model}

So far we have focused on the quark structures of the proton. However, interesting measurements have also been reported recently for the neutron charge radius~\cite{Atac:2021wqj}, which could provide an additional nontrivial constraint on the understanding of the quark structures of nucleons. In the $\Delta$ and $Y$ models above, the flavor structures and the spatial wave functions are ``factorized'' (i.e., independent of each other). If this assumption were the full story, then the neutron charge radius would vanish, which would be in contradiction with the experimentally suggested nonzero squared charge radius $(-0.110 \pm 0.008)\rm fm^2$ from~\cite{Atac:2021wqj}. 

 {
It is most plausible that what entails a nonzero squared charge radius for neutron is the dynamically generated nontrivial correlations among both constituent and sea quarks/antiquarks. One well known mechanism manifesting such correlations is through virtual pionic clouds that play  a key role in understanding the electromangetic form factors of nucleons (especially that of the neutron)~\cite{Miller:2002ig,Cloet:2012cy,Glozman:1998yh,Hammer:2003qv}.  
}

To account for the charge radius of both proton and neutron together in the present potential model, one has to go beyond the naive assumption of factorized flavor and spatial structures. 
 {It is not obvious how the aforementioned pionic effects could be simply introduced into our model, so we resort to the idea of diquark correlations/clustering within hadrons~\cite{Jaffe:1975fd,Wilczek:2004im,Carlson:2005pe,Lebed:2015tna,Barabanov2021,Ferretti2019}. This mechanism is straightforward to implement and is a reasonable approach, given that the main goal of this study is to investigate the difference between charge and mass distributions within a given model.} 
In such a quark-diquark model for nucleons, a $u$ quark and a $d$ quark tend to have strong spatial correlations in the color-$\bar{\mathbf{3}}$, Lorentz-scalar and iso-scalar channel due to nonperturbative QCD interactions (e.g., arising from instantons~\cite{Schafer:1996wv}). In this section, we explore the implication of such a diquark model with the assumption of a tightly bound $u$-$d$ diquark (as an effective quasi-particle) which can be bound to either an additional $u$ quark to form a proton or an additional $d$ quark to form a neutron.

A nucleon consisting of one quark and one point-like $u$-$d$ diquark  is effectively a two-body system. This has been solved for the Cornell Hamiltonian using a variety of methods\cite{Hall1984,Hall2015,Plante2005}. We solve the  Schr\"odinger equation  with the potential in eq. (\ref{eq_Cornell}) using the method described by Fulcher \cite{Fulcher1993} in the center-of-mass frame: 
\begin{equation}
    \left(
    \frac{\left(\alpha+1\right)^2}{\alpha}\mu
    +\frac{p^2}{2\mu}
    -\frac{\kappa}{x}
    +\sigma x
    -V_0
    \right)
    \psi
    =M\psi,
\end{equation}
where $\bm{x}\equiv\bm{r}_2-\bm{r}_1$ and $\bm{p}$ is the conjugate momentum; $\mu\equiv m_1m_2/(m_1+m_2)$ is the reduced mass and $\alpha\equiv m_1/m_2$ is the diquark to quark mass ratio. Note there are two mass parameters in this model: $m_1$ for the diquark mass and $m_2$ for the single quark. We use the same potential parameters $\kappa$ and $\sigma$ as before while fix the offset $V_0$ from the nucleon mass. We use the $s$-wave basis functions
\begin{equation}
    \psi_n(x)
    =A_ne^{-\beta x}L_n^2(2\beta x),
    \quad
    n=0,1,\cdots,
\end{equation}
with normalization
\begin{equation}
    A_n^2
    =\frac{2\beta^3}{\pi(n+1)(n+2)},
\end{equation}
where $L_n^2$ is an associated Laguerre polynomial. The parameter $\beta$ is a scale factor that affects the rate of convergence. These basis functions satisfy the orthogonality relation
\begin{equation}
    \int
    d^3x\,
    \psi_n^*(x)
    \psi_{n'}(x)
    =\delta_{nn'}.
\end{equation}
The kinetic matrix elements are
\begin{equation}
    \bra{n}\frac{p^2}{2\mu}\ket{n'}
    =\frac{\beta^2}{\mu}
    \sqrt{\frac{(n+1)(n+2)}{(n'+1)(n'+2)}}
    \left(
    1+\frac{2n}{3}-\frac{1}{2}\delta_{nn'}
    \right),
    \quad
    n\leq n';
\end{equation}
the potential matrix elements are
\begin{equation}
    \bra{n}\frac{\kappa}{r}\ket{n'}
    =\kappa\beta
    \sqrt{\frac{(n+1)(n+2)}{(n'+1)(n'+2)}},
    \quad
    n\leq n',
\end{equation}
and
\begin{equation}
    \bra{n}\sigma r\ket{n'}
    =\frac{\sigma}{2\beta}
    \left(
    (2n+3)\delta_{nn'}
    -\sqrt{n'(n'+2)}\delta_{n',n+1}
    -\sqrt{n(n+2)}\delta_{n',n-1}
    \right).
\end{equation}
It is sufficient to set $n=20$ and $\beta=1\,\text{GeV}$ for the system at hand. Then, through diagonalization of the Hamiltonian, one can obtain the ground state energy and the corresponding wave function. Note that, in this model, the proton and neutron are both described by the ground state wave function with degenerate energy, and the only difference between them is the flavor content of the single quark.

\begin{figure}[!htb]
    \centering
    \includegraphics[width=0.6\textwidth]{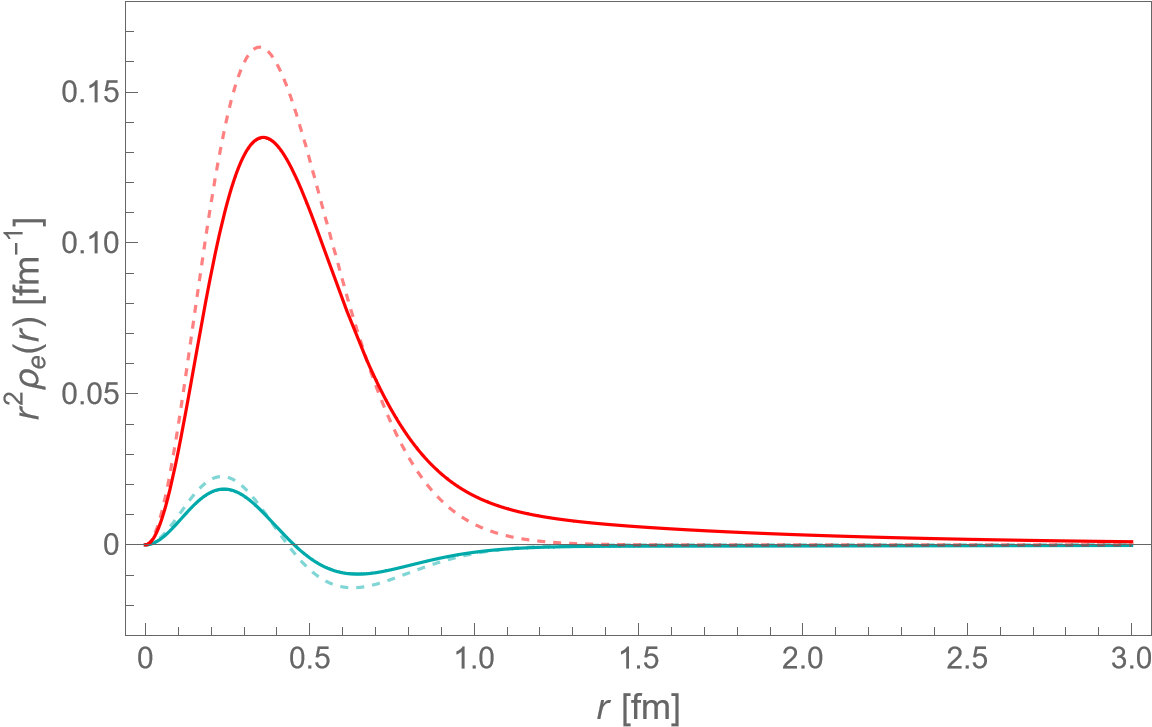}
    \caption{Proton (red) and neutron (cyan) charge distributions. The substitution $\delta\rightarrow D_i$ smears the charge densities, increasing the proton charge radius and the magnitude of the squared neutron charge radius. The dashed curves correspond to point-like quarks/diquarks and the solid curves incorporate smearing.}
    \label{fig_rho_e_diquark}
\end{figure}

\begin{figure}[!htb]
    \centering
    \includegraphics[width=0.5\textwidth]{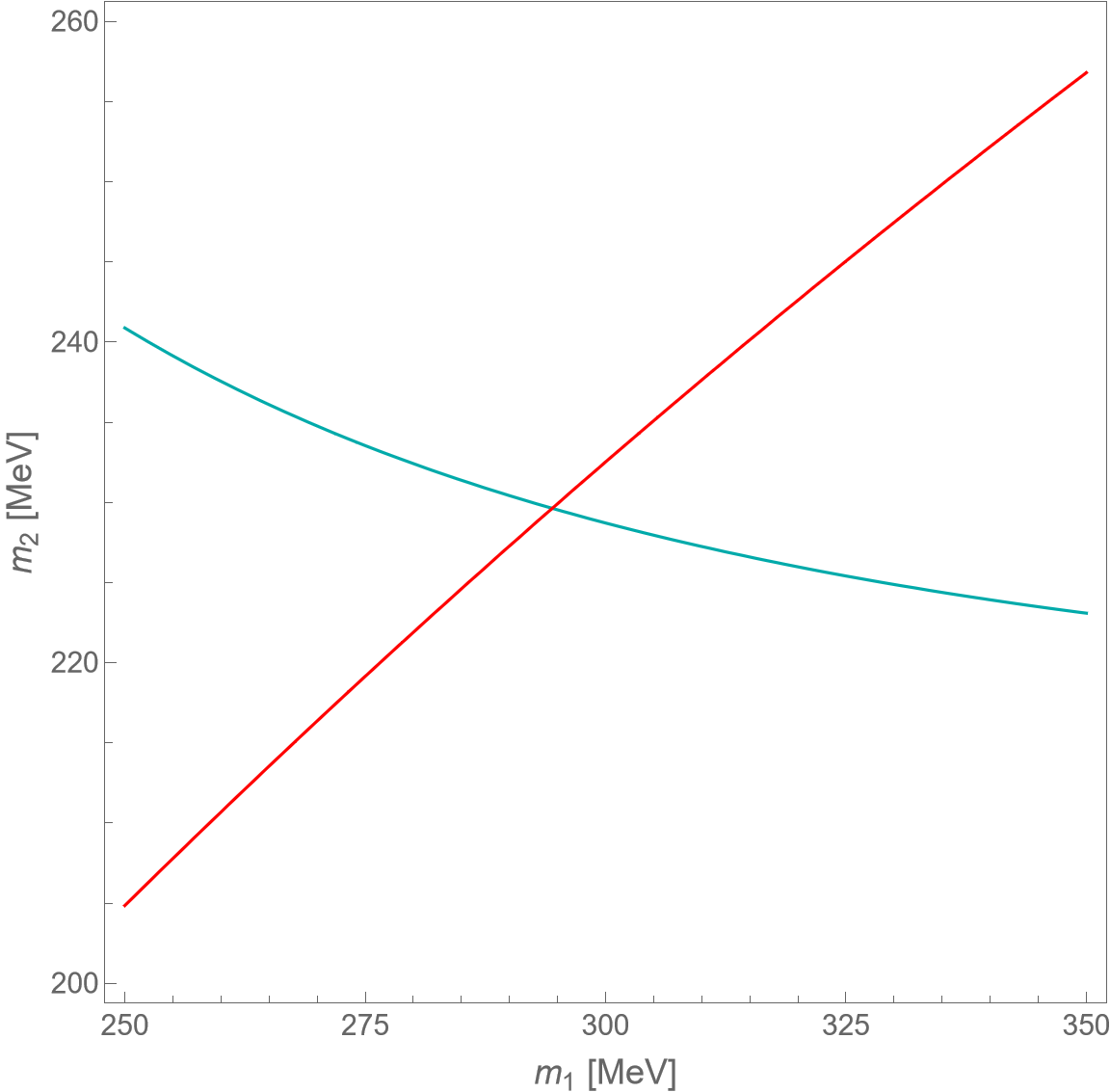}
    \caption{Diquark and single quark mass parameters $m_1$ and $m_2$ that give the experimental values of the  proton (red) and neutron (cyan) squared charge radii, respectively. The two curves intersect  at $m_1=294\,\text{MeV}$, $m_2=230\,\text{MeV}$.}
    \label{fig_diquark_curves}
\end{figure}

With the above solution, we can then calculate the charge distributions of both the proton and neutron in the diquark model in a similar manner to eq. (\ref{eq_rho_e}), albeit only summing over a $u$-$d$ diquark with electric charge $1/3$ and a $u$ quark with charge $2/3$ for the proton or a $d$ quark with charge $-1/3$ for the neutron. Examples of such charge distributions are shown in fig.~ \ref{fig_rho_e_diquark}, from which the squared charge radii can be readily computed and compared with experimental values. Fig.~\ref{fig_diquark_curves} demonstrates that masses $m_1$ and $m_2$ exist such that both the squared proton and neutron charge radii agree with experiment. The red curve indicates $(m_1,m_2)$ combinations that correctly give the proton squared charge radius while the cyan curve indicates $(m_1,m_2)$ combinations that correctly give the proton squared charge radius. The intersection of the two curves at $m_1=294\,\text{MeV}$ and $m_2=230\,\text{MeV}$ is the optimal pair of masses. We notice that while the single quark mass is reasonably close to typical constituent quark mass values in the literature, the required diquark mass appears rather light. This could be due to the specific assumption of the present model with a tightly-bound point-like diquark. One could imagine that this becomes different in other models with different implementations of diquark correlations.   






Finally, we can compute the mass distribution of a nucleon in the diquark model (noting that this shall be identical for both the proton and neutron) in a way similar to that in the $\Delta$ or $Y$ models. In this case, the kinetic terms are 
\begin{equation}
    \braket{\tau_i}
    =\frac{1}{2m_i}
    \int d^3x
    \,D_i(\bm{r}-\bm{r}_i)
    |\bm{\nabla}_{r_i}\psi|^2,
\end{equation}
with $i=1$ representing the diquark and $i=2$ the remaining single quark. The integral can be written entirely in terms of the internal quantities $\bm{x}$ and $\bm{p}$ using the relationships
\begin{align}
    \bm{r}_1
    &=-\frac{m_2\bm{x}}{m_1+m_2}, \\
    \bm{r}_2
    &=\frac{m_1\bm{x}}{m_1+m_2},
\end{align}
and
\begin{equation}
    \bm{p}_{r_1}
    =-\bm{p}_{r_2}
    =-\bm{p}.
\end{equation}
The potential energy from the Coulomb and linear terms will be treated in the same way as before. Again, a key component is the offset energy, determined to be $V_0=664.5\,\text{MeV}$ in the diquark model and spatially distributed according to eq.~\ref{eq_offset_dist}.

\begin{figure}[!htb]
    \centering
    \includegraphics[width=0.6\textwidth]{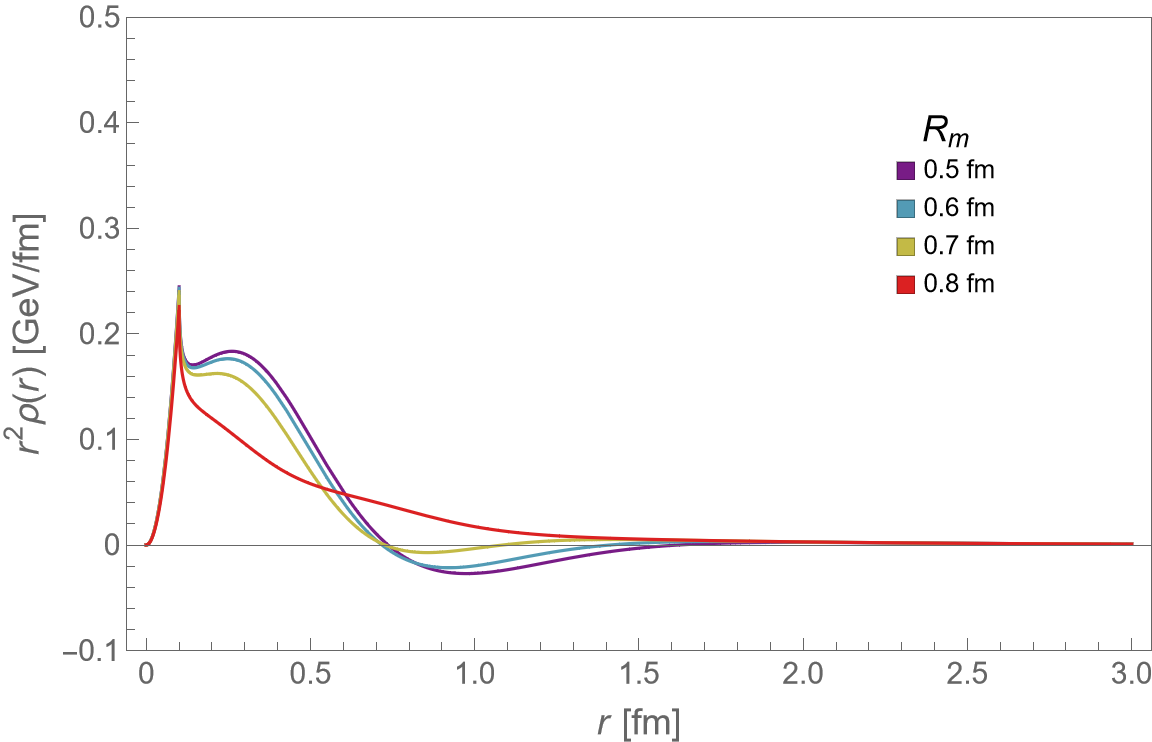}
    \caption{Nucleon mass distributions in the diquark model corresponding to different choices of mass radius $R_m$.}
    \label{fig_massrad_diquark}
\end{figure}

In fig.~\ref{fig_massrad_diquark}, we display the nucleon mass distribution obtained from the diquark model for several choices of the width parameter $w$, which corresponds to a range of different mass radii. The relationship between the mass radius $R_m$ and the width parameter $w$ from the diquark model is already shown as the dashed curve in  fig.~\ref{fig_radius_vs_width}. Compared with that from the $\Delta$ and $Y$ models, the mass radius from the diquark model demonstrates a visibly more ``flattened'' dependence on the width parameter. However, it is interesting to note that the three different models appear to converge toward one another in the vicinity around $R_m \simeq (0.5\sim 0.6)\rm fm$, which happens to be an area favored by a number of recent analyses. We end this section by emphasizing the constraining power of considering the squared charge radii of both proton and neutron together and the advantage of the diquark model (or similar models with suitable flavor correlations) in providing a simultaneous description of them.    

\section{Implications for Initial Conditions in Heavy Ion Collisions}

In collisions between nuclei, the initial geometry of the collision determines the evolution of the collision products. This is captured by the Glauber model, which was developed to address high energy scattering of composite particles. For large nuclei, the optical limit of the Glauber model gives analytical expressions for many observables of interest \cite{Miller2007}. However, because it assumes continuous projectile and target nucleon densities, it is unable to capture event-by-event local density fluctuations. In contrast, the Monte Carlo implementation of the Glauber model represents each nucleon individually and maintains good agreement with experiment for a wider variety of initial conditions.

At sufficiently high energies and short time scales, nuclei behave as collections of uncorrelated nucleons traveling on straight-line trajectories. In the Monte Carlo Glauber model, the nucleons are sampled from a Woods-Saxon distribution,
\begin{equation}
    \rho_\text{WS}(r,\theta)
    =\frac{\rho_0}{1+\exp\left(a^{-1}\left(r-R\left(1+\beta_2Y_{2,0}(\theta)+\beta_4Y_{4,0}(\theta)\right)\right)\right)}.
\end{equation}
The parameters $a$ and $R$ are the skin depth and nuclear radius, respectively; $\beta_1$ and $\beta_2$ encode deviations from a spherically symmetric density. Hard-core repulsion between nucleons in the rest frame of a given nucleus is enforced by requiring the centers of any pair of nucleons to be separated by a minimum distance ($0.9\,\text{fm}$ in this study). Due to the ultrarelativistic energies involved, each nucleus is contracted into a pancake along the beam direction, so that the collision takes place in a plane transverse to the beam. Projectile and target nucleons collide if their transverse separation is less than $\sqrt{\sigma_{NN}/\pi}$, where $\sigma_{NN}$ is the inelastic nucleon-nucleon cross-section (black disc model). The colliding nucleons, called participants, may experience many sequential binary collisions. 

The geometry of the collision region can be characterized using eccentricities. The $n$th \textit{participant} eccentricity is defined
\begin{equation}
    \varepsilon_n
    =\left\lvert
    \frac{\int dx\,dy\,s(x,y)r_\perp^ne^{in\phi}}{\int dx\,dy\,s(x,y)r_\perp^n}
    \right\rvert,
\end{equation}
where $r_\perp^2=x^2+y^2$ and $\tan\phi=y/x$, and the integrals are weighted by the transverse entropy density $s(x,y)$. For $n=2$, this is equivalent to
\begin{equation}
    \varepsilon_2
    =\frac{\sqrt{(\sigma_x^2-\sigma_y^2)^2+4(\sigma_{xy}^2)^2}}{\sigma_x^2+\sigma_y^2}.
\end{equation}
With enough nucleons, the collision products evolve hydrodynamically in accord with the geometry of the impact region. The $n$th eccentricity $\varepsilon_n$ is proportional to the $n$th Fourier coefficient $v_n$ of the azimuthal asymmetry; e.g., the ellipticity $\varepsilon_2$ is proportional to the elliptic flow $v_2$.

\begin{figure}[!htb]
    \centering
    \includegraphics[width=0.6\textwidth]{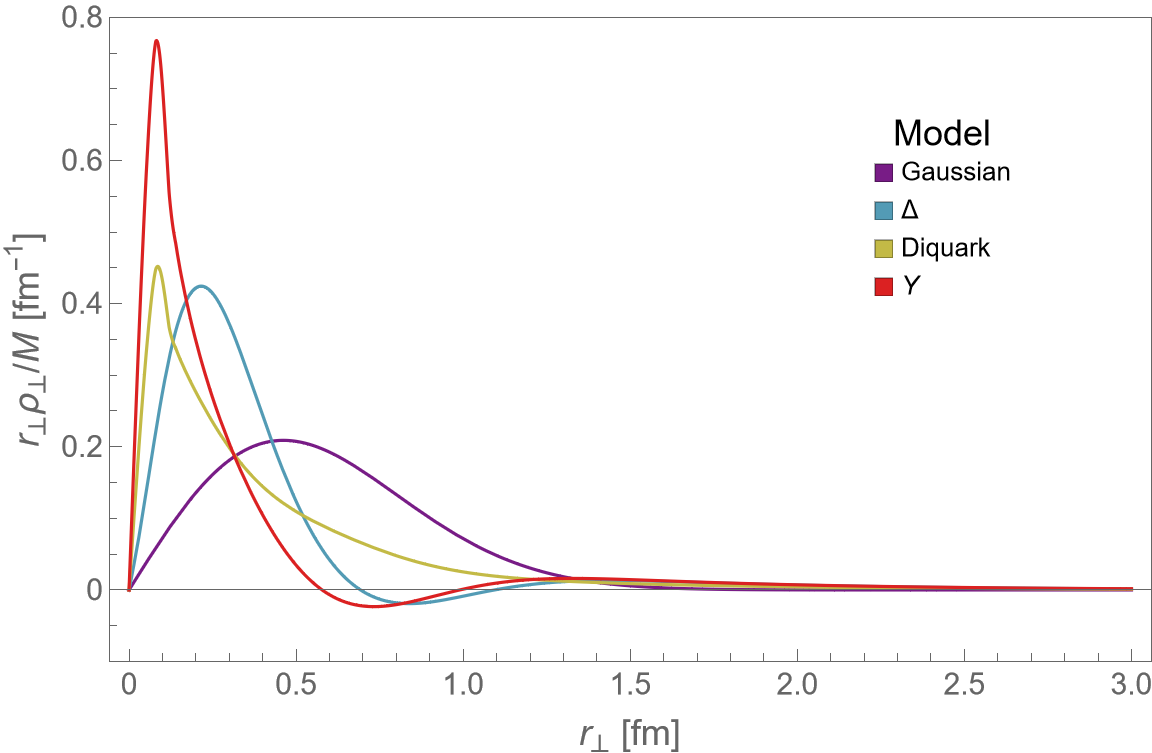}
    \caption{The transverse mass distribution $\rho_\perp$ is the total mass distribution $\rho$ integrated over the $z$-coordinate; $r_\perp$ is the transverse radius defined by $r^2=z^2+r_\perp^2$. Pictured here are nucleon profiles with rest frame RMS radius $0.8\,\text{fm}$.}
    \label{fig_transverse}
\end{figure}

The entropy density 
\begin{equation}
    s(\bm{x})
    \propto
    \frac{1-\alpha_\text{glb}}{2}
    \rho_\text{part}(\bm{x})
    +\alpha_\text{glb}
    \rho_\text{coll}(\bm{x})
\end{equation}
is computed from the binary collision density
\begin{equation}
    \rho_\text{coll}(\bm{x})
    =\sum_{i\in\text{coll}}
    w_\text{coll}^i
    p_\perp(\bm{x}-\bm{x}_i)
\end{equation}
and the participant density
\begin{equation}
    \rho_\text{part}(\bm{x})
    =\sum_{i\in\text{part}}
    w_\text{part}^i
    p_\perp(\bm{x}-\bm{x}_i)
\end{equation}
with mixing parameter $\alpha_\text{glb}=0.118$, where $\bm{x}=(x,y)$ and $\bm{x}_i$ is the location of a collision or participant. The weights $w_\text{coll}^i$ and $w_\text{part}^i$ follow a gamma distribution with a mean of unity, allowing fluctuations in multiplicity. The normalized distribution $p_\perp$ is the transverse nucleon profile. The standard choice is
\begin{equation}
    p_\perp(\bm{x}-\bm{x}_i)
    =\frac{1}{2\pi w_N^2}
    \exp\left(-\frac{\left(\bm{x}-\bm{x}_i\right)^2}{2w_N^2}\right),
\end{equation}
with the RMS radius in the rest frame equal to $\sqrt{3}w_N$. We simulate collisions using this profile as a baseline, in addition to the transverse mass densities of the $\Delta$, diquark, and $Y$ models normalized by total mass, i.e., $\rho_\perp/M$ (see fig. \ref{fig_transverse}).

The center-of-mass collision energies used in this study are $\sqrt{s_{NN}}=5.02\,\text{TeV}$ for Pb+Pb and p+Pb, $7\,\text{TeV}$ for O+O, and $9.9\,\text{TeV}$ for p+O. The energies $7\,\text{TeV}$ and $9.9\,\text{TeV}$ have been proposed in future O+O and p+O experiments as part of the LHC RUN3 \cite{Behera2022,Adriani2019}. For each pair of nuclei, data was gathered after $10^5$ events since statistical errors in the simulation at this level are small compared to the uncertainty in real-world collision experiments at LHC energies. Also, because the alternative nucleon profiles in fig. \ref{fig_transverse} were calculated numerically from
\begin{equation}
    p_\perp(r_\perp)
    =\frac{1}{M}
    \int_{-\infty}^\infty dz\,\rho\!\left(\sqrt{z^2+r_\perp^2}\right),
\end{equation}
they do not have compact formulas. Instead, the above integral is evaluated for $|z|<8\,\text{fm}$ from $r_\perp=0$ to $5\,\text{fm}$ every $0.02\,\text{fm}$, for a total of $251$ points representing the curve $p_\perp$. This list of points was imported into the collision code, and whenever a value of $p_\perp$ was needed that was not in the list, a straight-line interpolation between points was used.
\begin{figure}[!htb]

\setcounter{subfig}{0}
  \newcommand\typecap{\stepcounter{subfig}\captiontext*[\value{subfig}]{}}
  \captionsetup{position=bottom, skip=3pt}
  \centering
  \begin{subcaptiongroup}
    \captionlistentry{}\label{fig_e2_Y_sensitivity}   
    \captionlistentry{}\label{fig_e3_Y_sensitivity}
    \tikz[
      node distance=3pt,
      caption/.style={
        anchor=north east,
        font=\Large,
        outer sep=10pt,
      },
      relpos/.style={
        right=of #1.south east,
        anchor=south west,
      },
    ] {
      \node (A) {\includegraphics[width=0.48\textwidth]{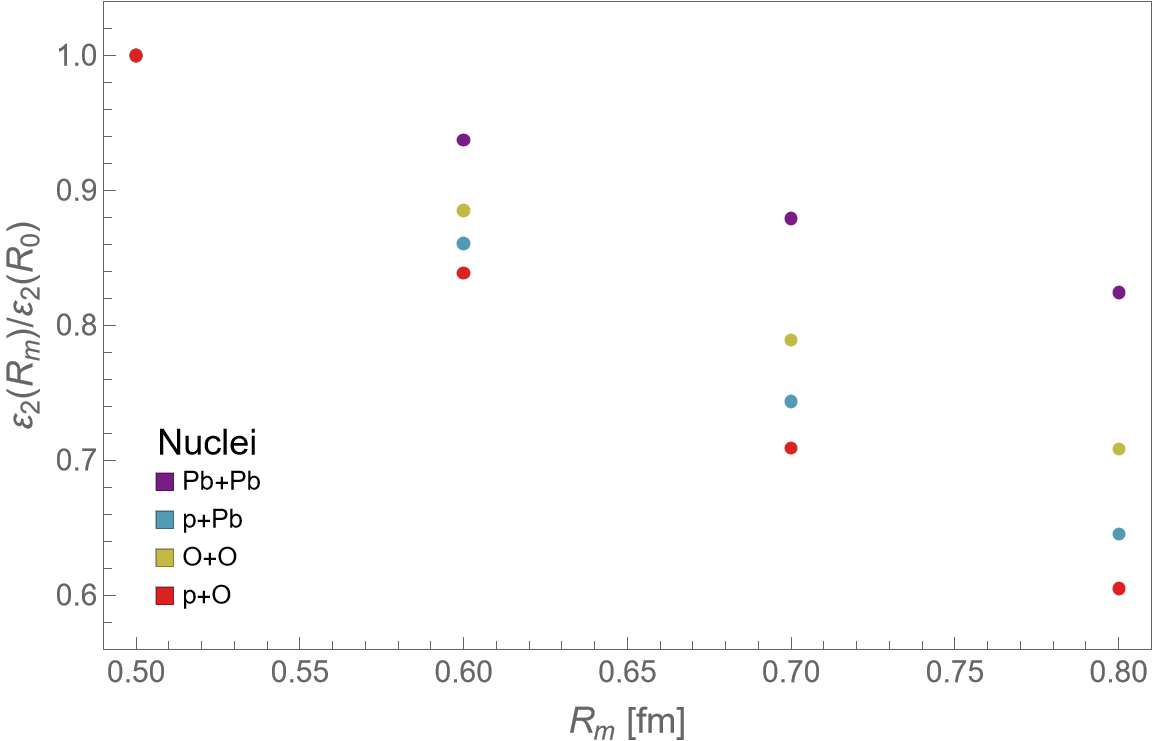}};
      \node (B) [relpos=A] {\includegraphics[width=0.48\textwidth]{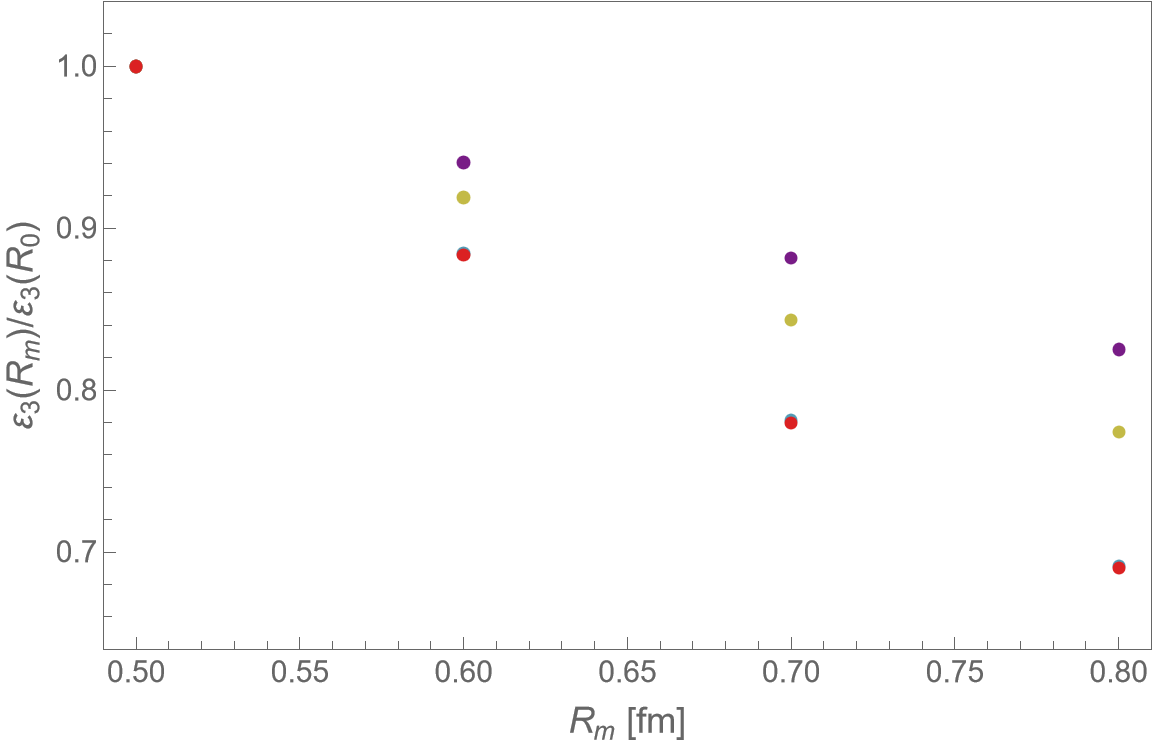}};
      \node at (A.north east) [caption] {\typecap};
      \node at (B.north east) [caption] {\typecap};
    }
  \end{subcaptiongroup}
  \caption{Ratios of ellipticity (\subref{fig_e2_Y_sensitivity}) and triangularity (\subref{fig_e3_Y_sensitivity}) in the $Y$ model with respect to baseline values at $R_0=0.5\,\text{fm}$. In (\subref{fig_e3_Y_sensitivity}), the p+Pb and p+O systems overlap. Impact parameters and center-of-mass energies for each collision system are given in the Appendix. Similar behaviors are observed also for other models of nucleon mass distributions, and detailed results are included in the Appendix.}
   \label{fig_Y_sensitivities}

\end{figure}


We find that the collision geometry is indeed sensitive to nucleon size, particularly in peripheral collisions or collisions with few participants ($N_\text{part}\lesssim30$). This can be seen in fig. \ref{fig_Y_sensitivities}, where the eccentricities of the p+Pb and p+O systems at $R_m=0.8\,\text{fm}$ are less than $70\%$ of the eccentricities at $R_m=0.5\,\text{fm}$. In general, a smaller radius results in a larger ellipticity and triangularity. It is worth noting that the sensitivity of these eccentricities to the mass radius $R_m$ are significant enough to allow a strategy of putting meaningful quantitative constraints on $R_m$ through high precision measurements of collective flows in those colliding systems.

While all models produce similar $\varepsilon_2$ for a given mass radius, the Gaussian nucleon results in a noticeably different $\varepsilon_3$. This is likely due to $\varepsilon_2$ capturing the overall shape of the collision, which is roughly the overlap of two discs, and $\varepsilon_3$ capturing fluctuations away from that simple geometry. That is to say, the nucleon profile does not have much effect on the overall shape of the collision region, which depends more on the positions of the centers of the participants. However, the profile will have an effect on fluctuations since each of the models---$\Delta$, diquark, and $Y$---produce nucleon profiles with a more pronounced core compared to a Gaussian, which can be seen in fig. \ref{fig_transverse}. The complete eccentricity vs. impact parameter curves (figs. \ref{fig_PbPb}--\ref{fig_pO}) and peripheral collision data (figs. \ref{fig_fracs_PbPb}--\ref{fig_fracs_pO}) are included in the Appendix.

\section{Conclusion}

In summary, we've presented a detailed study of the charge and mass distributions within a nucleon (both proton and neutron included) and the associated squared radii based on a potential model approach. A number of different constituent quark configurations have been considered and compared. Possible implications for the initial conditions in heavy ion collisions have been explored. 

We conclude this work by emphasizing a few key findings below. First, while the charge radius is dictated by quark dynamics, the mass radius is strongly influenced by nonperturbative QCD contributions to a nucleon's mass that are not sensitive to the constituent quarks. This important feature allows the possibility of a substantial difference between the charge radius and the mass radius. Second, a simultaneous description of the radii of both proton and neutron could provide valuable constraints on the quark structures of nucleon, in particular on the flavor-sensitive correlations such as diquark clustering. Finally, simulations demonstrate a considerable sensitivity of the eccentricities in the initial conditions of heavy ion collisions to the input nucleon profiles, in particular to the corresponding overall size characterized by the mass radius. 

All of these taken together would suggest that high precision measurements from both hadronic reactions and heavy ion collisions can together provide an accurate extraction of nucleon's mass radius which could offer unique insights on its mass component  from nonperturbative QCD dynamics.




\section*{Acknowledgement}

The authors thank A. Akridge, X.R. Chen, D. Kharzeev, M. Shepherd and A. Szczepaniak for useful discussions. This work is partly supported by the U.S. NSF under Grant No.~PHY-2209183 and by the U.S. DOE through ExoHad Topical Collaboration. Gallimore  acknowledges financial support from The Gordon and Betty Moore Foundation and the American Physical Society to present preliminary results of this work at the GHP 2023 workshop.

\section*{Appendix}

\begin{figure}[!h]
    \centering
    \includegraphics[width=\textwidth]{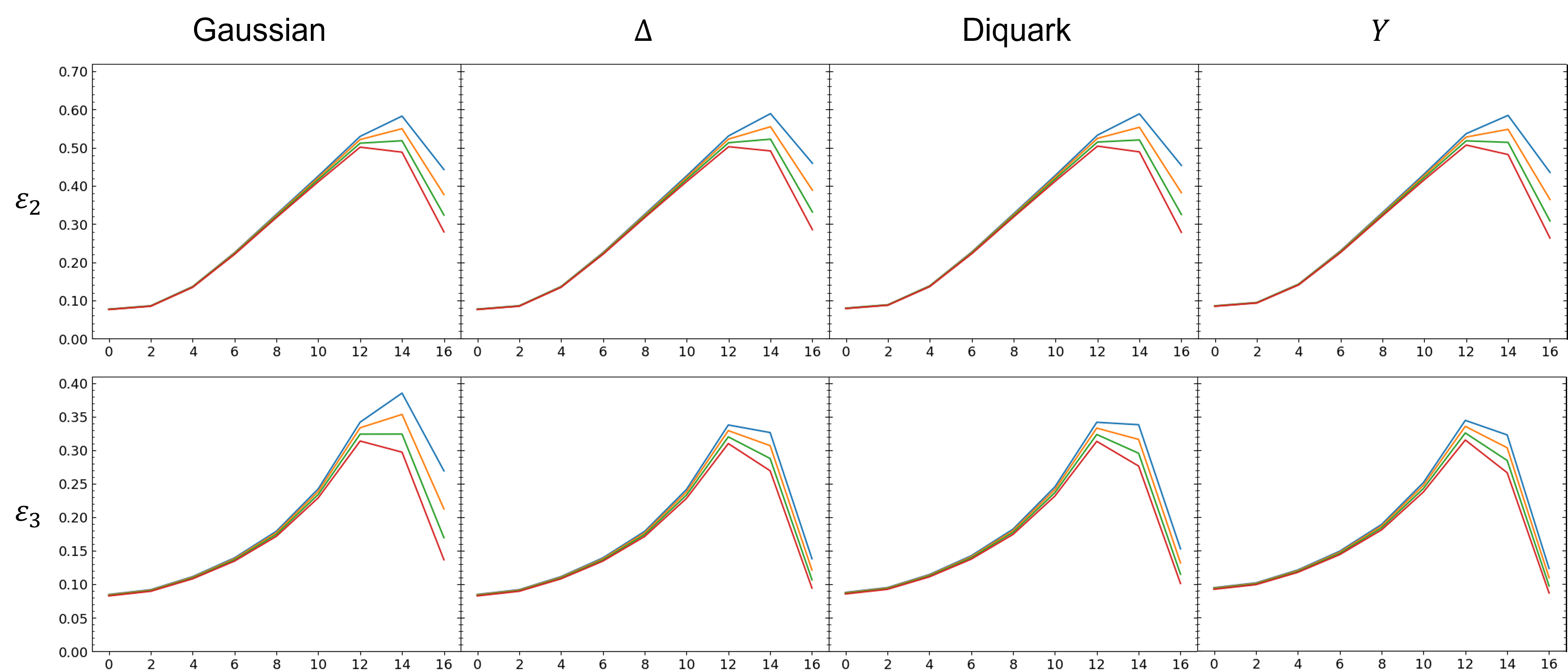}
    \caption{Pb+Pb collisions at $\sqrt{s_{NN}}=5.02\,\text{TeV}$ for different observables (rows) and nucleon profile models (columns). Horizontal axis corresponds to impact parameter $b$. Curves represent mass radii $0.5\,\text{fm}$ (blue), $0.6\,\text{fm}$ (orange), $0.7\,\text{fm}$ (green), and $0.8\,\text{fm}$ (red).}
    \label{fig_PbPb}
\end{figure}

\begin{figure}[!h]
    \centering
    \includegraphics[width=\textwidth]{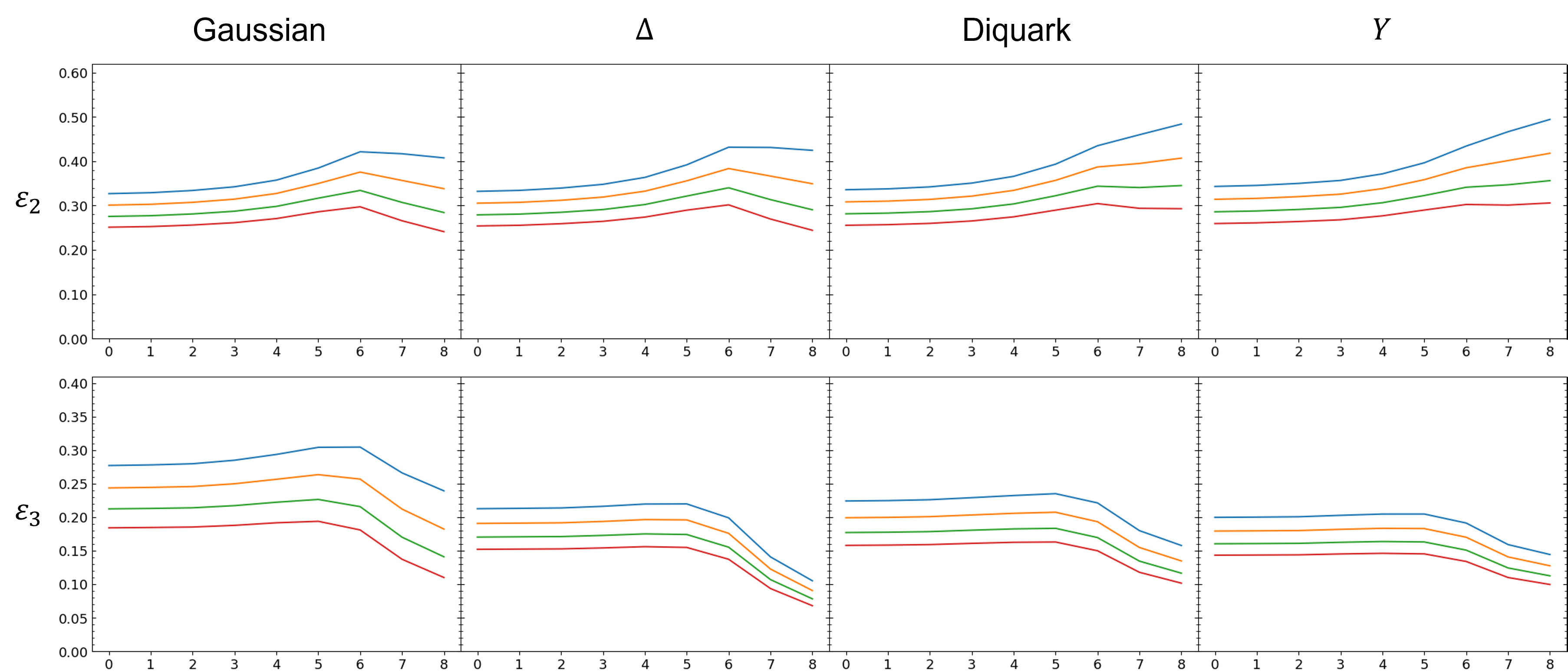}
    \caption{p+Pb collisions at $\sqrt{s_{NN}}=5.02\,\text{TeV}$ for different observables (rows) and nucleon profile models (columns). Horizontal axis corresponds to impact parameter $b$. Curves represent mass radii $0.5\,\text{fm}$ (blue), $0.6\,\text{fm}$ (orange), $0.7\,\text{fm}$ (green), and $0.8\,\text{fm}$ (red).}
    \label{fig_pPb}
\end{figure}

\begin{figure}[!h]
    \centering
    \includegraphics[width=\textwidth]{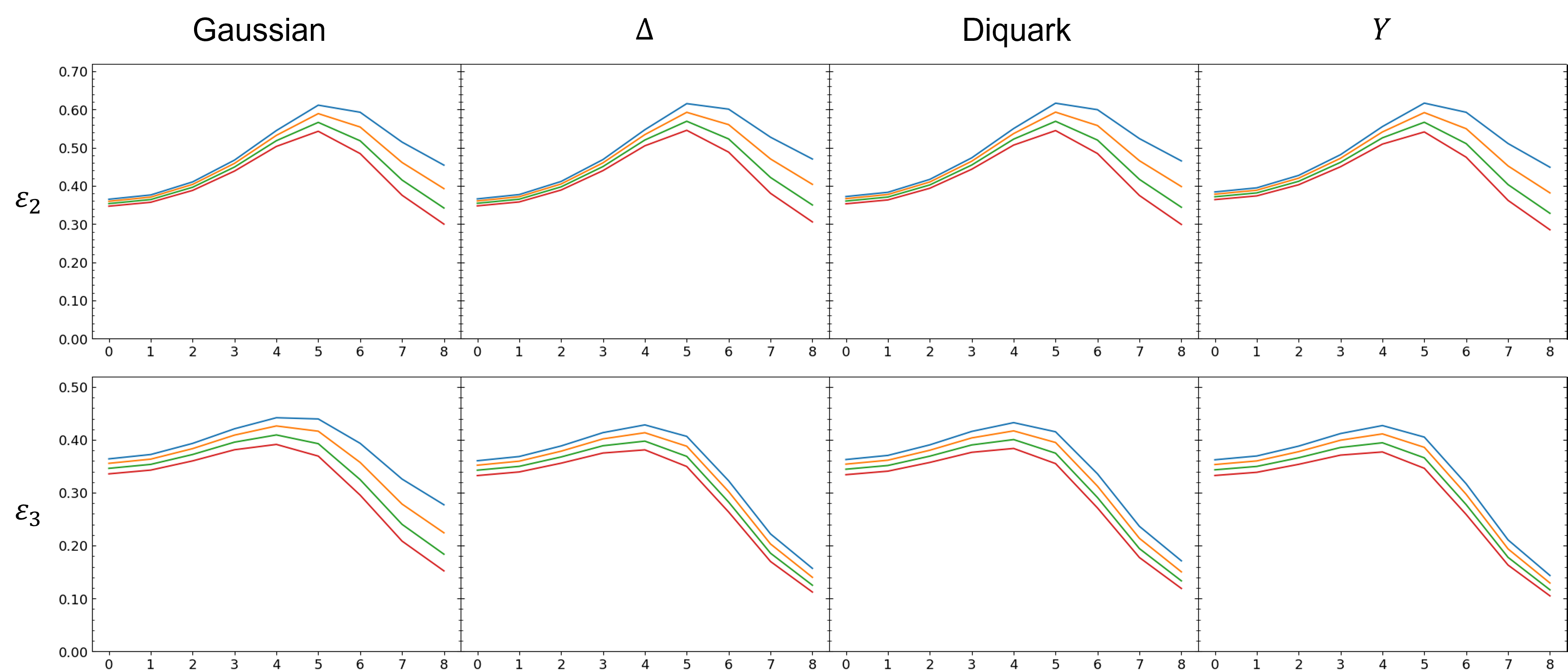}
    \caption{O+O collisions at $\sqrt{s_{NN}}=7\,\text{TeV}$ for different observables (rows) and nucleon profile models (columns). Horizontal axis corresponds to impact parameter $b$. Curves represent mass radii $0.5\,\text{fm}$ (blue), $0.6\,\text{fm}$ (orange), $0.7\,\text{fm}$ (green), and $0.8\,\text{fm}$ (red).}
    \label{fig_OO}
\end{figure}

\begin{figure}[!h]
    \centering
    \includegraphics[width=\textwidth]{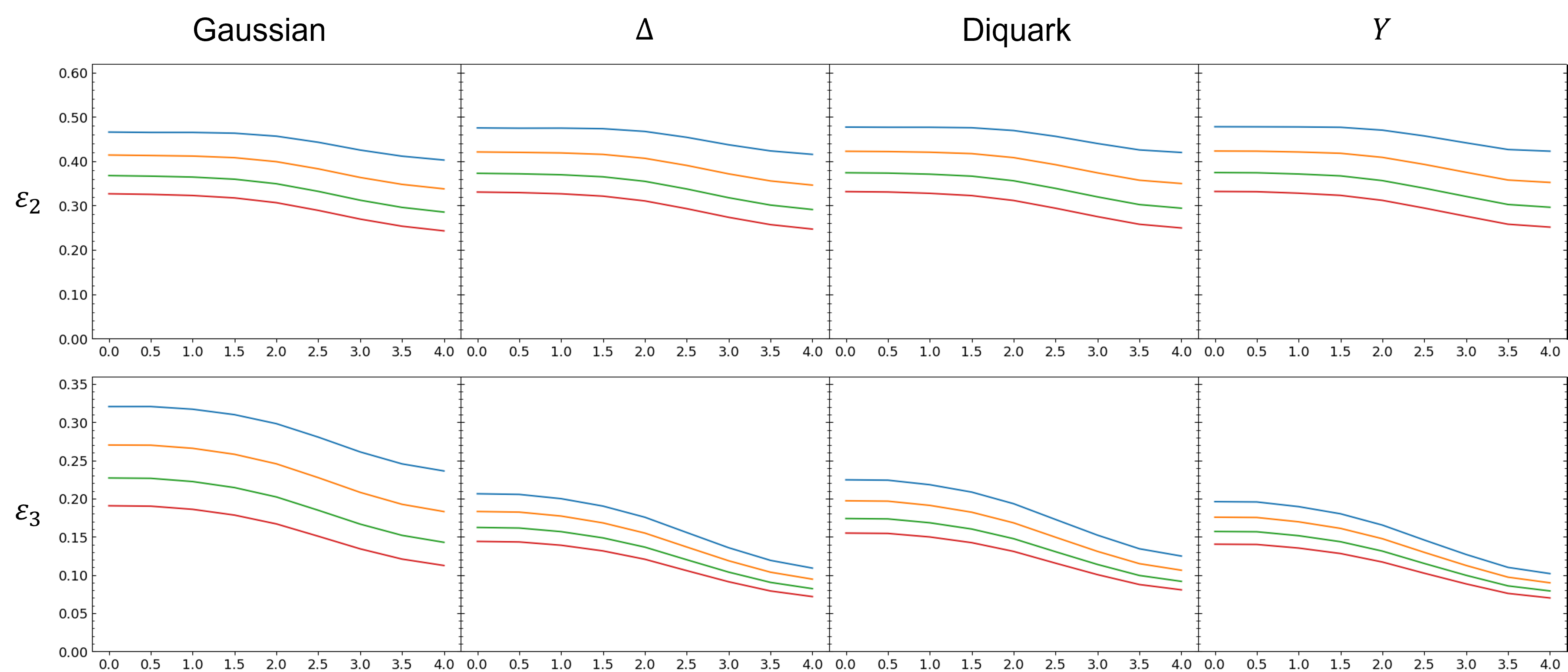}
    \caption{p+O collisions at $\sqrt{s_{NN}}=9.9\,\text{TeV}$ for different observables (rows) and nucleon profile models (columns). Horizontal axis corresponds to impact parameter $b$. Curves represent mass radii $0.5\,\text{fm}$ (blue), $0.6\,\text{fm}$ (orange), $0.7\,\text{fm}$ (green), and $0.8\,\text{fm}$ (red).}
    \label{fig_pO}
\end{figure}

\begin{figure}[H]

\setcounter{subfig}{0}
  \newcommand\typecap{\stepcounter{subfig}\captiontext*[\value{subfig}]{}}
  \captionsetup{position=bottom, skip=3pt}
  \centering
  \begin{subcaptiongroup}
    \captionlistentry{}\label{fig_e2_frac_PbPb}   
    \captionlistentry{}\label{fig_e3_frac_PbPb}
    \tikz[
      node distance=3pt,
      caption/.style={
        anchor=north east,
        font=\Large,
        outer sep=10pt,
      },
      relpos/.style={
        right=of #1.south east,
        anchor=south west,
      },
    ] {
      \node (A) {\includegraphics[width=0.48\textwidth]{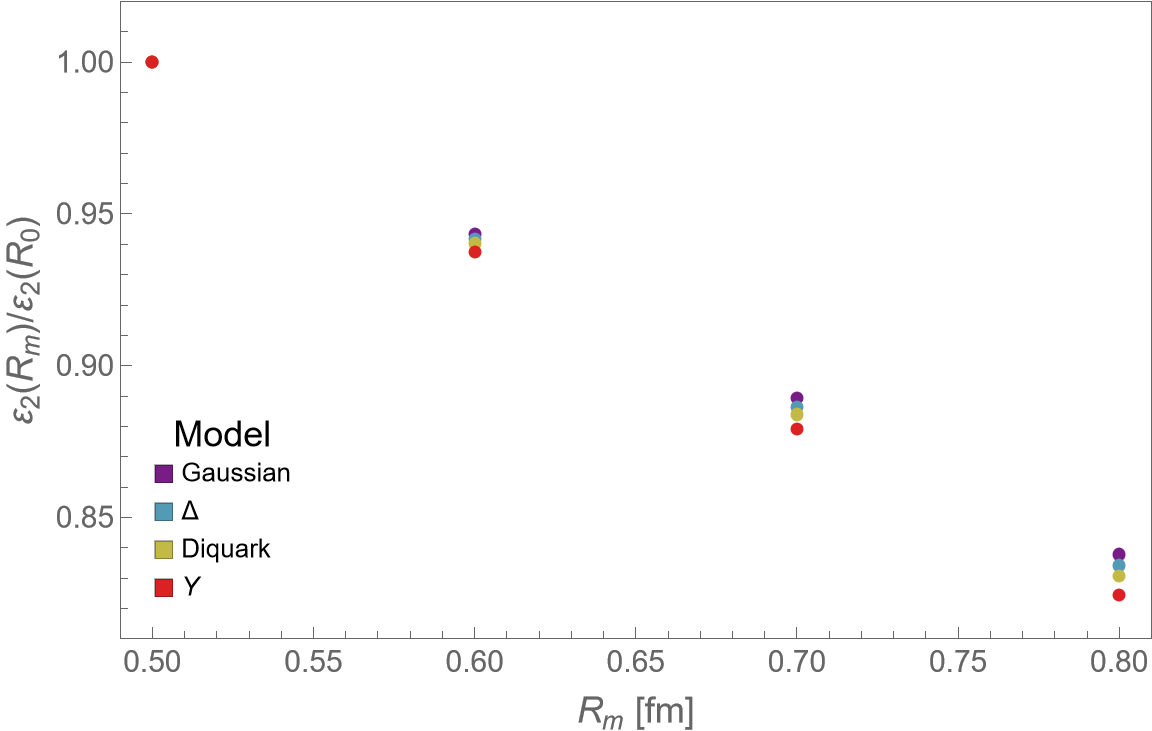}};
      \node (B) [relpos=A] {\includegraphics[width=0.48\textwidth]{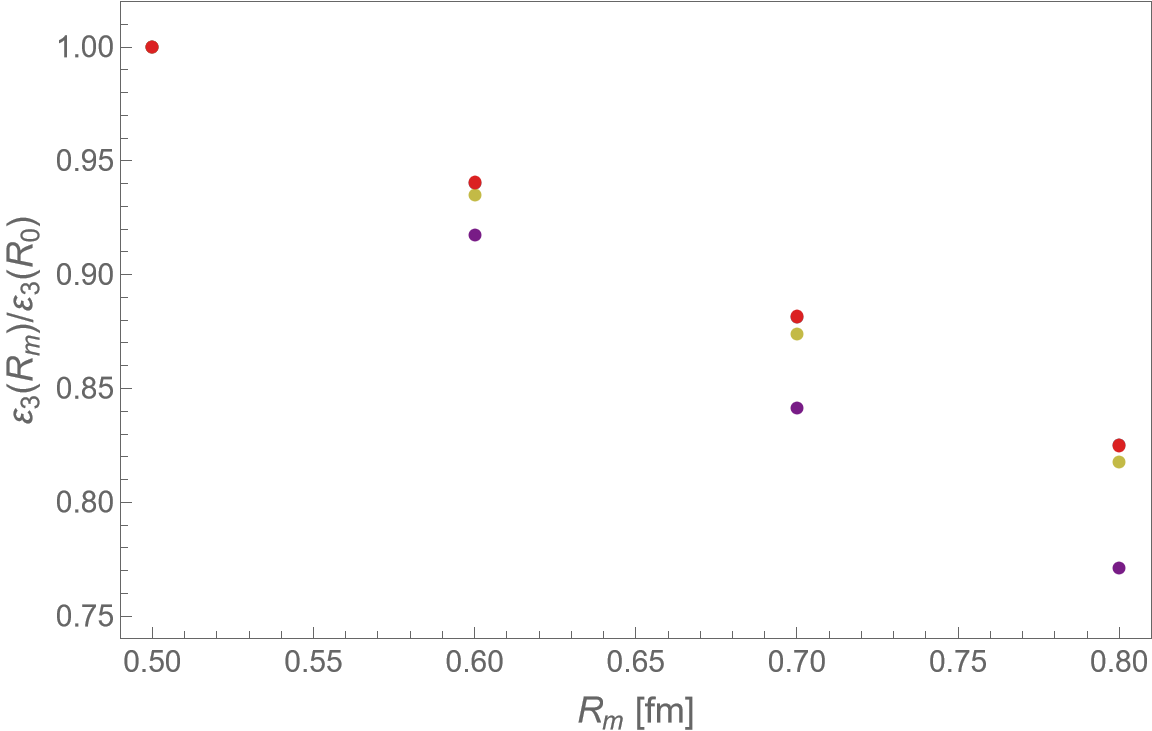}};
      \node at (A.north east) [caption] {\typecap};
      \node at (B.north east) [caption] {\typecap};
    }
  \end{subcaptiongroup}
  \caption{Ratios of ellipticity (\subref{fig_e2_frac_PbPb}) and triangularity (\subref{fig_e3_frac_PbPb}) with respect to baseline values at $R_0=0.5\,\text{fm}$ for Pb+Pb collisions at $\sqrt{s_{NN}}=5.02\,\text{TeV}$ and $b=14\,\text{fm}$.}
   \label{fig_fracs_PbPb}

\end{figure}

\begin{figure}[H]

\setcounter{subfig}{0}
  \newcommand\typecap{\stepcounter{subfig}\captiontext*[\value{subfig}]{}}
  \captionsetup{position=bottom, skip=3pt}
  \centering
  \begin{subcaptiongroup}
    \captionlistentry{}\label{fig_e2_frac_pPb}   
    \captionlistentry{}\label{fig_e3_frac_pPb}
    \tikz[
      node distance=3pt,
      caption/.style={
        anchor=north east,
        font=\Large,
        outer sep=10pt,
      },
      relpos/.style={
        right=of #1.south east,
        anchor=south west,
      },
    ] {
      \node (A) {\includegraphics[width=0.48\textwidth]{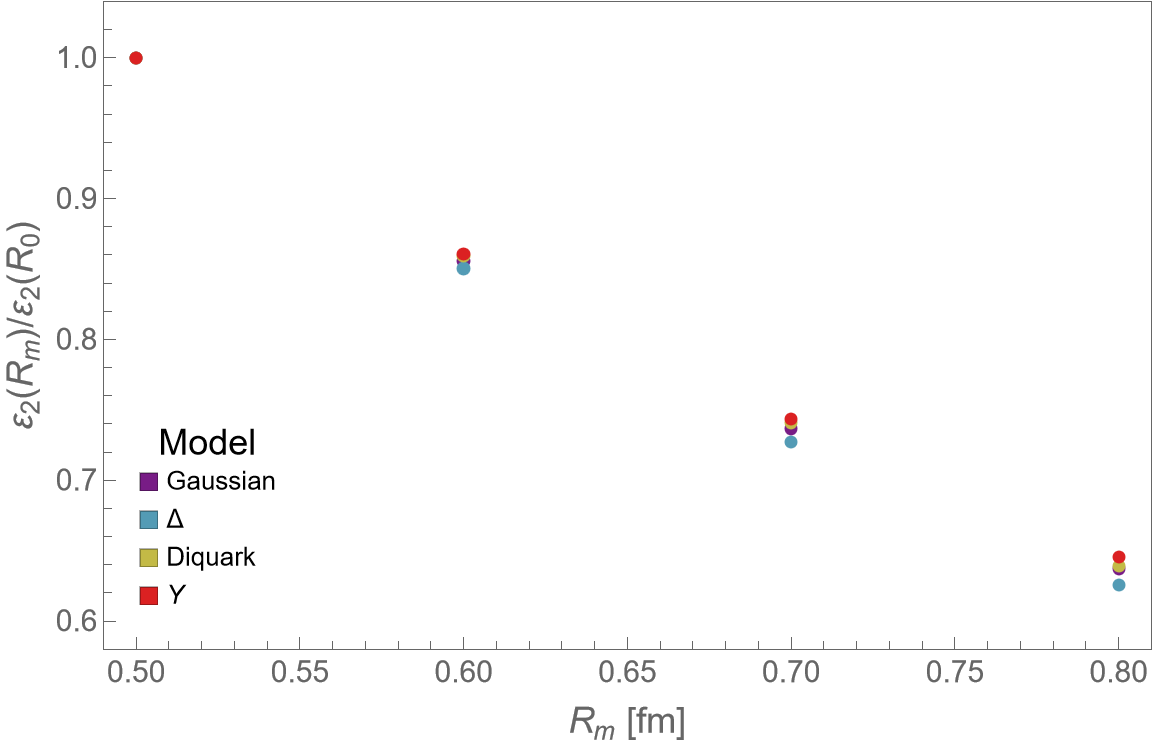}};
      \node (B) [relpos=A] {\includegraphics[width=0.48\textwidth]{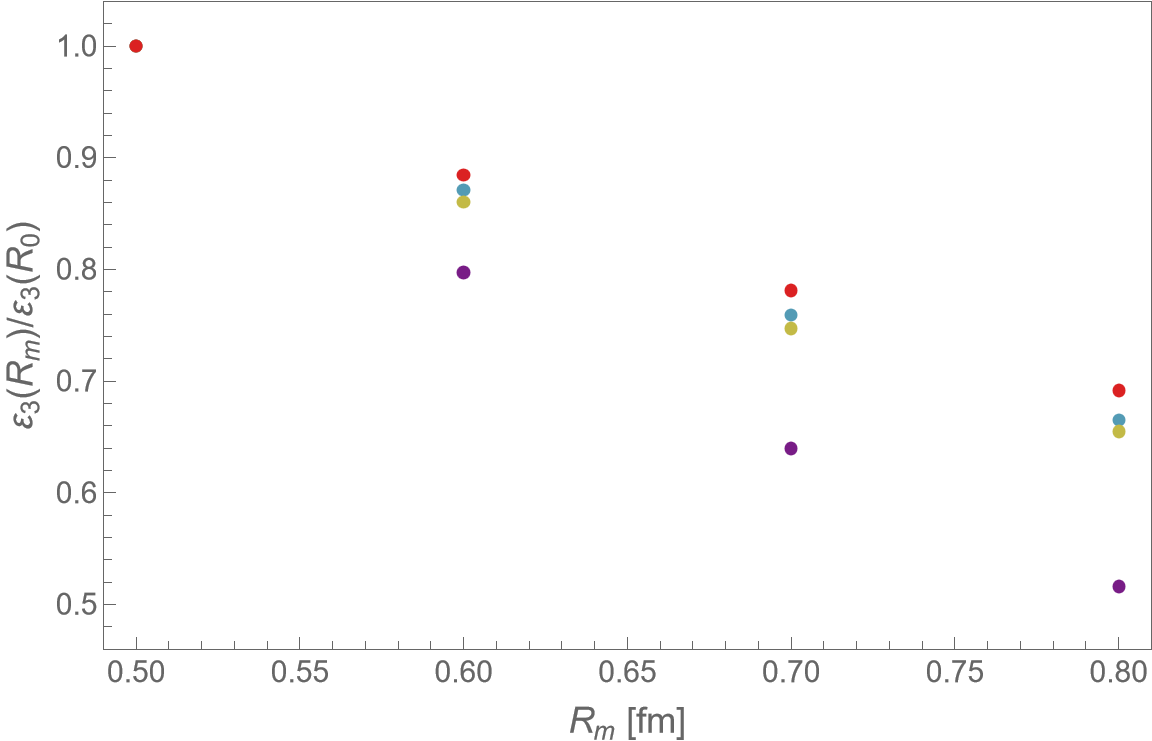}};
      \node at (A.north east) [caption] {\typecap};
      \node at (B.north east) [caption] {\typecap};
    }
  \end{subcaptiongroup}
  \caption{Ratios of ellipticity (\subref{fig_e2_frac_pPb}) and triangularity (\subref{fig_e3_frac_pPb}) with respect to baseline values at $R_0=0.5\,\text{fm}$ for p+Pb collisions at $\sqrt{s_{NN}}=5.02\,\text{TeV}$ and $b=7\,\text{fm}$.}
   \label{fig_fracs_pPb}

\end{figure}

\begin{figure}[H]

\setcounter{subfig}{0}
  \newcommand\typecap{\stepcounter{subfig}\captiontext*[\value{subfig}]{}}
  \captionsetup{position=bottom, skip=3pt}
  \centering
  \begin{subcaptiongroup}
    \captionlistentry{}\label{fig_e2_frac_OO}   
    \captionlistentry{}\label{fig_e3_frac_OO}
    \tikz[
      node distance=3pt,
      caption/.style={
        anchor=north east,
        font=\Large,
        outer sep=10pt,
      },
      relpos/.style={
        right=of #1.south east,
        anchor=south west,
      },
    ] {
      \node (A) {\includegraphics[width=0.48\textwidth]{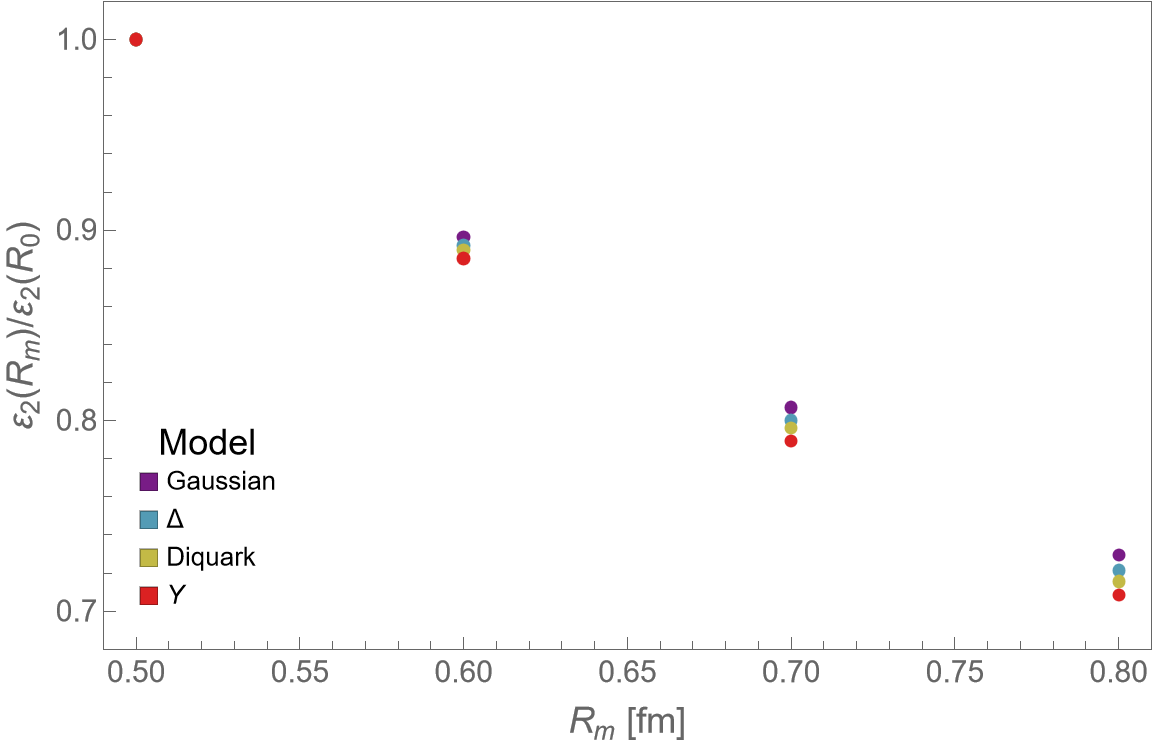}};
      \node (B) [relpos=A] {\includegraphics[width=0.48\textwidth]{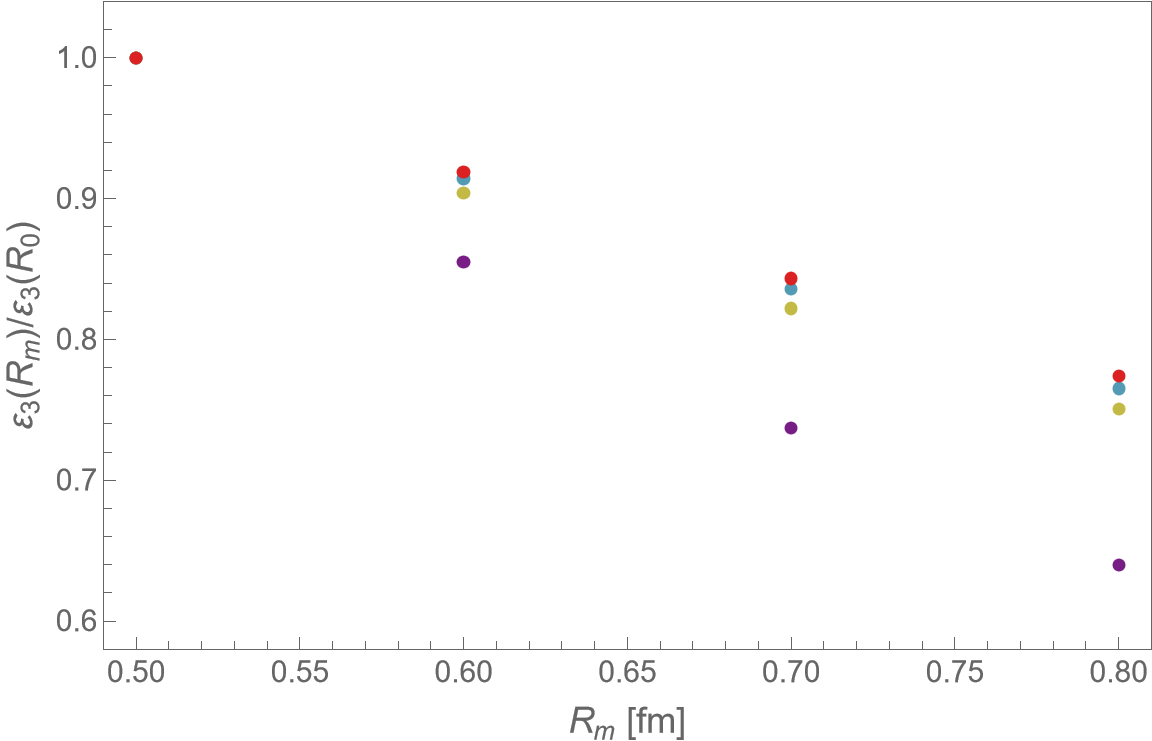}};
      \node at (A.north east) [caption] {\typecap};
      \node at (B.north east) [caption] {\typecap};
    }
  \end{subcaptiongroup}
  \caption{Ratios of ellipticity (\subref{fig_e2_frac_OO}) and triangularity (\subref{fig_e3_frac_OO}) with respect to baseline values at $R_0=0.5\,\text{fm}$ for O+O collisions at $\sqrt{s_{NN}}=7\,\text{TeV}$ and $b=7\,\text{fm}$.}
   \label{fig_fracs_OO}

\end{figure}

\begin{figure}[H]

\setcounter{subfig}{0}
  \newcommand\typecap{\stepcounter{subfig}\captiontext*[\value{subfig}]{}}
  \captionsetup{position=bottom, skip=3pt}
  \centering
  \begin{subcaptiongroup}
    \captionlistentry{}\label{fig_e2_frac_pO}   
    \captionlistentry{}\label{fig_e3_frac_pO}
    \tikz[
      node distance=3pt,
      caption/.style={
        anchor=north east,
        font=\Large,
        outer sep=10pt,
      },
      relpos/.style={
        right=of #1.south east,
        anchor=south west,
      },
    ] {
      \node (A) {\includegraphics[width=0.48\textwidth]{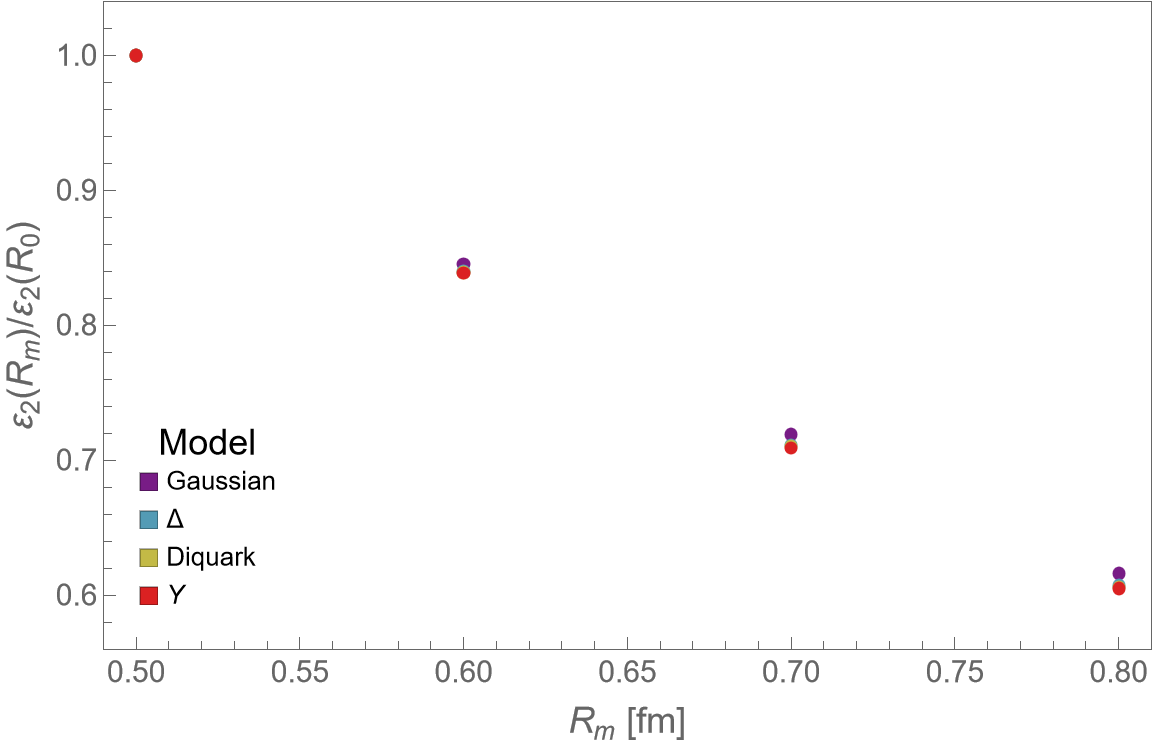}};
      \node (B) [relpos=A] {\includegraphics[width=0.48\textwidth]{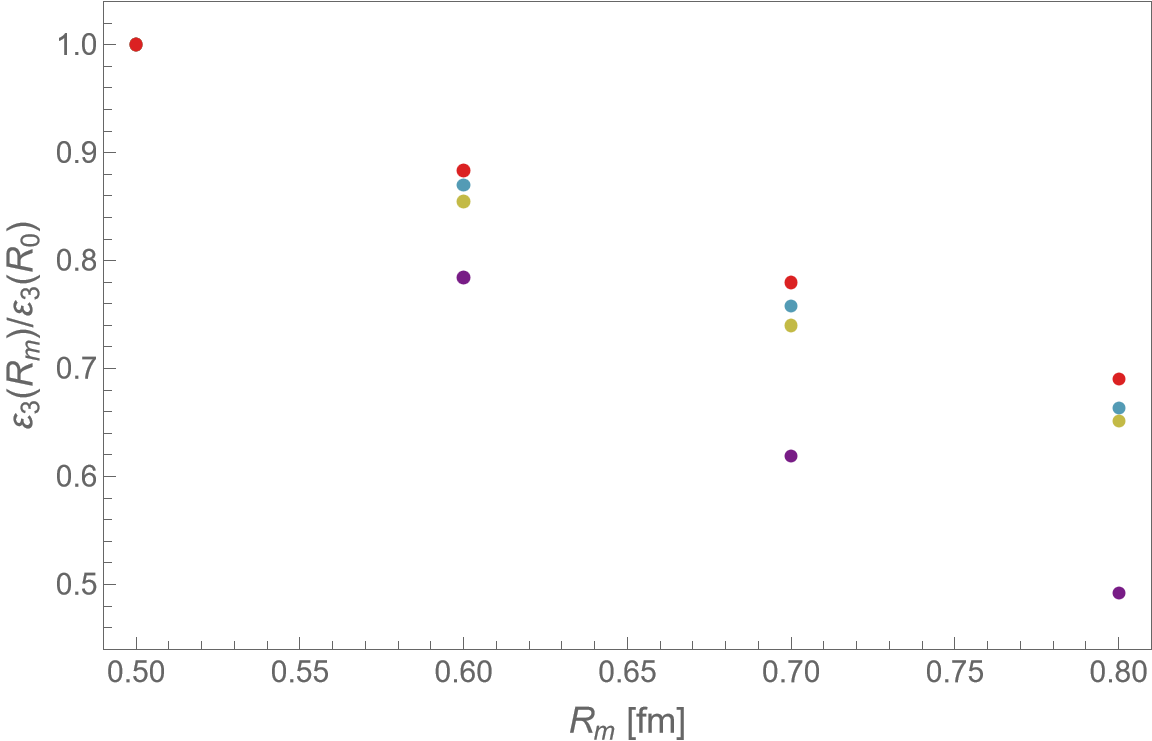}};
      \node at (A.north east) [caption] {\typecap};
      \node at (B.north east) [caption] {\typecap};
    }
  \end{subcaptiongroup}
  \caption{Ratios of ellipticity (\subref{fig_e2_frac_pO}) and triangularity (\subref{fig_e3_frac_pO}) with respect to baseline values at $R_0=0.5\,\text{fm}$ for p+O collisions at $\sqrt{s_{NN}}=9.9\,\text{TeV}$ and $b=3.5\,\text{fm}$.}
   \label{fig_fracs_pO}

\end{figure}

Included below (figs. \ref{fig_PbPb}--\ref{fig_pO}) are the eccentricities of the initial collision geometry for different collision systems. For a given system, the different models behave similarly, with a smaller mass radius producing a larger $\varepsilon_2$ or $\varepsilon_3$. Special attention is given to the peripheral region, where differences between mass radii tend to be more pronounced, in figs. \ref{fig_fracs_PbPb}--\ref{fig_fracs_pO}.

\newpage

\bibliography{MassRad.bib}

\end{document}